\documentclass[10pt,a4paper,twocolumn,floatfix]{extarticle}
\usepackage{cab_arxiv_style}

\usepackage[numbers,sort&compress,round]{natbib}
\newcommand{\SI}[0]{\textcolor{CBgreyblue}{Supplementary Information}}
\newcommand{\SIapp}[1]{\textcolor{CBgreyblue}{Supplementary Information, section #1}}
\newcommand{\SIFig}[1]{\textcolor{CBgreyblue}{Supplementary Fig.~S{#1}}}
\newcommand{\methods}[1]{Methods}

\usepackage[colorlinks,
  pdfpagelabels=false,
  hyperfootnotes=false,
  citecolor=CBgreyblue,
  hyperindex=false,
  urlcolor=blue,
  linkcolor=CBgreyblue,
  bookmarks=false,
  hypertexnames=true,
  linktoc=false,
  hyperindex=false,
]{hyperref}

\title{Simulating the chromatin mediated phase separation of model proteins with multiple domains}

\author{Marco Ancona}
\author{Chris A. Brackley} 
\affil{SUPA, School of Physics and Astronomy, University of Edinburgh, Peter Guthrie Tait Road, Edinburgh EH9
 3FD, United Kingdom}

\renewcommand{\keywords}{Keywords: chromatin $|$ phase separation $|$ bridging-induced attraction $|$ polymer simulation}

\begin{document}

\twocolumn[
  \begin{@twocolumnfalse}
\maketitle

\thispagestyle{fancy}

\begin{abstract}
We perform simulations of a system containing simple model proteins and a polymer representing chromatin. We study the interplay between protein-protein and protein-chromatin interactions, and the resulting condensates which arise due to liquid-liquid phase separation, or a via a `bridging-induced attraction' mechanism. For proteins which interact multivalently, we obtain a phase diagram which includes liquid-like droplets, droplets with absorbed polymer, and coated polymer regimes. Of particular interest is a regime where protein droplets only form due to interaction with the polymer; here, unlike a standard phase separating system, droplet density rather than size varies with the overall protein concentration. We also observe that protein dynamics within droplets slow down as chromatin is absorbed. If the protein-protein interactions have a strictly limited valence, fractal or gel-like condensates are instead observed. Together this provides biologically relevant insights into the nature of protein-chromatin condensates in living cells. 
\end{abstract}

\clearpage


  \end{@twocolumnfalse}
]

\thispagestyle{fancy}

The cell nucleus is a highly structured organelle which contains much of an organism's genetic material~\cite{Alberts2014}. This material exists as chromatin, a composite of DNA and histone proteins which makes up the chromosomes. While the nucleus itself is surrounded by a membrane, most of the structures within it [known as `nuclear bodies'~\cite{Mao2011}] are membraneless assemblies of proteins, DNA and/or RNA. Some of these, including nucleoli, Cajal bodies, and splicing speckles, are found in the interchromatin regions. Others co-localise with chromatin, and examples include: clusters of transcription factors, RNA polymerase II, and other proteins associated with transcription~\cite{Sabari2018,Chong2018}; polycomb bodies, involved in cell-type specific gene repression~\cite{Bantignies2011,Eeftens2020}; and foci of heterochromatin, a tightly packaged form of chromatin which tends to be transcriptionally repressed~\cite{Brero2005,Probst2008}. 

There has been much recent interest in how protein foci form in the nucleus, and whether a liquid-liquid phase separation (LLPS) mechanism plays a role. A common notion is that flexible, low complexity and intrinsically disordered protein (IDP) domains facilitate LLPS~\cite{Brangwynne2015}. IDPs often contain exposed charges or hydrophobic residues, leading to weak multivalent attractive interactions; having multiple interaction points and a `coil' configuration is thought to lead to interactions which are effectively longer ranged than those between globular proteins~\cite{Martin2020}. Many IDPs, and several proteins which possess both disordered and globular domains, have indeed been found to readily phase separate \textit{in vitro}. 

Another mechanism which can (in the presence of chromatin) lead to protein phase separation, is the ``bridging-induced attraction'' (BIA). This was first uncovered in simulations studying how protein-chromatin interactions can drive chromosome organisation~\cite{Brackley2013,Brackley2016}, and was more recently demonstrated \textit{in vitro}~\cite{Ryu2021}. It arises when proteins or protein complexes with multiple DNA/chromatin binding domains form molecular bridges between different chromatin regions. The first protein to form a bridge produces a local increase in chromatin density, which leads to further protein binding and bridging at that location; this positive feedback ultimately gives rise to protein clustering. For the case of proteins which bind non-specifically to any chromatin site, the clusters will grow and coarsen until a single protein-rich phase remains~\cite{Brackley2017}; when there is an excess of proteins this also leads to chromatin compaction~\cite{Barbieri2012}.
Importantly, the BIA can give rise to phase separated foci in the absence of protein-protein interactions; we call this bridging-induced phase separation (BIPS). In Ref.~\cite{Brackley2020JPCM} it was shown that for model proteins with a finite number of chromatin binding domains, the shape of the protein can determine its ability to form bridges: for proteins which readily form bridges (``good bridgers''), the BIA is in effect and there is strong clustering and compaction. For poor bridgers, the BIA is not (or is only weakly) in effect, and protein clustering is not observed. 

The idea that LLPS is involved in genome regulation gained popularity after it was shown that heterochromatin protein 1 (HP1), one of the chief constituents of heterochromatin, was found to undergo phase separation \textit{in vitro}~\cite{Larson2017,Strom2017}. HP1 is highly conserved in eukaryotes, and is known to co-localise with heterochromatin foci~\cite{Eissenberg2000}. Its exact function in heterochromatin formation and gene silencing, however, remains elusive; possibilities are that it directly drives chromatin compaction, that it sterically occludes binding of activating proteins, or that it recruits further gene silencing machinery~\cite{Allshire2018,Sanulli2019}. In mammals there are three paralogs: HP1$\alpha$ and HP1$\beta$ are thought to have distinct roles in heterochromatin function, while HP1$\gamma$ also has a function in active chromatin~\cite{Canzio2014}. All have a similar structure, with two globular domains and three flexible/disordered regions. 
In the nucleus, HP1 is mainly found in dimers~\cite{Canzio2014,Sanulli2019} which have two chromatin binding domains, and so these can in principal form bridges. 

In this paper we study the interplay between LLPS and BIPS, considering how they could drive protein-chromatin foci localisation and compaction \textit{in vivo}. Inspired by work on patchy particles~\cite{Teixeira2017,Bianchi2008,Zaccarelli2007,Fusco2013,Zhang2004}, we have developed a simple coarse-grained model protein which resembles HP1, and we simulate these in solution with a chromatin fibre. More specifically, we consider two separate models which mimic two microscopic possibilities: (i) that the low-complexity domains give rise to weak and longer-ranged multivalent protein-protein attractions; and (ii) that the interactions between flexible domains are short ranged and have a limited valence such that exactly two domains can interact at a time. The first case involves a scenario where the flexible domains adopt an extended coil configuration, meaning multiple coils can overlap and there will be multiple weakly interacting contact points. The second case could arise, for example, when a disordered protein domain forms a globular secondary structure when interacting with the correct binding partner~\cite{Sugase2007}.
We explore the parameter space of the two systems in order to understand under what conditions aggregates containing both proteins and chromatin form, and measure the structural and dynamical properties. Importantly, our scheme is simple enough to allow us to perform simulations at many different points in parameter space, but retains details of the domain structure of the protein (explicitly incorporating protein-protein and protein-DNA interaction domains). Although the model is inspired by HP1, due to its simplicity we expect our results to be applicable more widely.

\begin{figure}[t]
\includegraphics{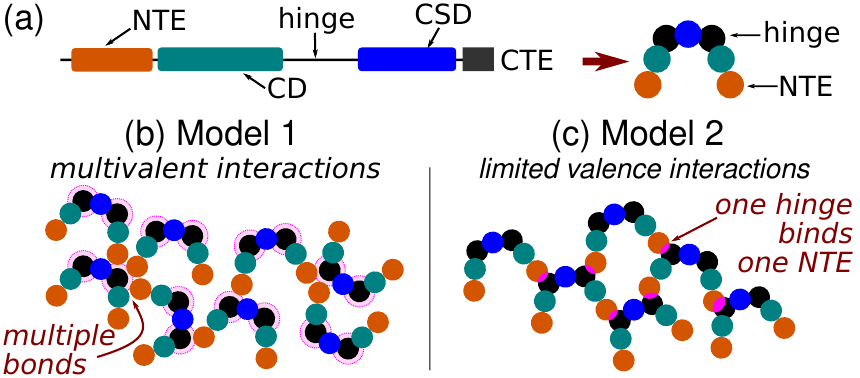}
\caption{\textbf{A simple coarse-grained protein model inspired by HP1.} (a) Left: schematic showing the domain structure of the HP1 protein as detailed in the text. Right: schematic representation of the model HP1 dimer. (b) and (c) Two alternative models for interactions between HP1 dimers. \label{fig:schematic}}
\end{figure}

\section*{Results}

\subsection*{Simulation scheme}

In this work we use coarse-grained Langevin dynamics simulations to study the behaviour of a system of simple HP1-inspired model proteins interacting with a model chromatin fibre.

\begin{figure*}[!ht]
\includegraphics{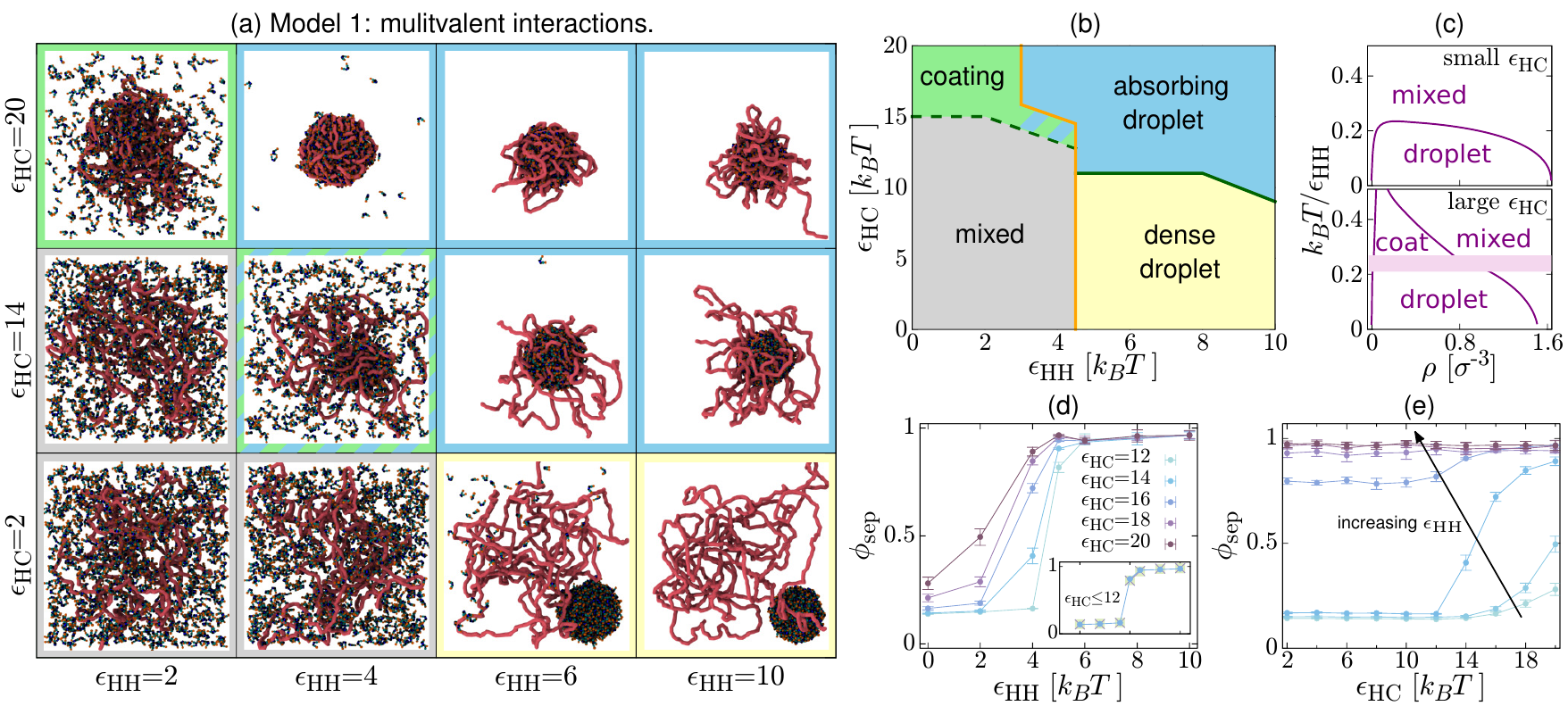}
\caption{\textbf{Protein-chromatin and multivalent protein-protein interactions lead to phase separation.}  (a) Snapshots of equilibrium configurations for $N\!=\!1000$ multivalent HP1 dimers interacting with an $L\!=\!1000$ bead polymer representing a 1~Mbp chromosome fragment for different values of the HP1-HP1 and HP1-chromatin interaction energies. (b) Phase diagram showing the different behaviours of the multivalent HP1s with different parameter values. Border colours in (a) indicate the relevant region in (b). (c) Phase diagram on the $\rho$-$\epsilon_{\rm HH}$ plane for small $\epsilon_{\rm HC}$ (top) and large $\epsilon_{\rm HC}$ (bottom). $\rho$ is the total number density of HP1s. In the bottom plot the shaded bar covers a region where there is a chromatin associated droplet, but the protein density inside and outside the droplet depends on the overall protein density (see text). These are sketch plots based on measurements of the HP1 density inside and outside of droplet [see \SIapp{6} and \SIFig{5}]. (d) Phase separation depth $\phi_{\rm sep}$ is plotted as a function of $\epsilon_{\rm HH}$ for different values of $\epsilon_{\rm HC}$ as indicated (units are $k_BT$). Each point is obtained from an average over 4 simulations of duration $5\!\times\!10^3\tau$ ($\tau$ is the simulation time unit, see \methods{}). Error bars show standard error in the mean; lines are a guide to the eye. The inset shows a similar plot for $\epsilon_{\rm HC}=$2,6,10 and 12$k_BT$ where points overlap. (e) $\phi_{\rm sep}$ is plotted as a function of $\epsilon_{\rm HC}$ for $\epsilon_{\rm HH}=0,2,4,5,6,8$ and $10k_BT$ (darker colours for larger values, as indicated by the arrow).\label{fig:iso_phase_diag}}
\end{figure*}

Figure~\ref{fig:schematic}(a) shows a schematic representation of the domains of HP1 [as detailed in, e.g., Refs.~\cite{Larson2017,Sanulli2019}]. These are known as the C-terminal end (CTE), the chromoshadow domain (CSD), a ``hinge'' region, the chromodomain (CD), and the N-terminal end (NTE). The CSD and CD are globular domains, while the others are flexible. Two HP1s form a dimer across the CSD, and the CD interacts with chromatin by binding tri-methylated lysines in the H3 histone (H3K9me3, a histone post-translational modification which is a hallmark of heterochromatin). In Ref.~\cite{Larson2017} it was shown that in human HP1, interaction between the hinge and the (phosphorylated) NTE allows further oligomerisation and, eventually, phase separation. 
Figure~\ref{fig:schematic}(a) also shows a schematic of how HP1 dimers (which we hereon refer to as HP1s) are represented in our simulations. They are modelled as rigid bodies made up from seven spheres, each representing a different domain (see \methods{} and \SIFig{1}). Our coarse-grained approach does not attempt to model the full details and exact dimensions of the dimer; nevertheless, we aim to capture the main features of the physics at the mesoscale. For chromatin we use a common coarse-grained polymer model where the fibre is represented as a chain of beads of diameter $\sigma\!=\!10$~nm connected by springs. The large-scale physical properties of the fibre are represented (i.e., its flexibility), but not the internal nucleosome structure. HP1 component spheres have a diameter 0.5$\sigma$; those representing CDs interact with polymer beads attractively, while all others interact with the polymer sterically only (see \methods{}). The dynamics of the polymer beads and HP1s (rigid body translation and rotation) are governed by a Langevin equation; we perform extensive simulations using the LAMMPS molecular dynamics software~\cite{Plimpton1995}. Details are given in \methods{} and \SI{}. 

As noted above, we study two versions of the model which differ in their protein-protein interactions. First, we consider \textit{multivalent interactions} [Fig.~\ref{fig:schematic}(b)], using a longer range interaction potential between the spheres representing the hinge and NTE, such that several NTEs can simultaneously interact with a hinge and \textit{vice versa} (determined by the geometry and steric hindrance).
Second, we consider \textit{limited valence interactions} [Fig.~\ref{fig:schematic}(c)], using a shorter ranged potential such that at most one hinge and one NTE can interact at a time.
Since an HP1 dimer has two hinges and two NTEs, in the limited valence model a given dimer can bind to at most four others at once. 
The strength of attractive interactions between HP1s, and between HP1s and chromatin are given by the energies $\epsilon_{\rm HH}$ and $\epsilon_{\rm HC}$ respectively. Since the interaction potentials differ, these two values should not be compared directly (nor should $\epsilon_{\rm HH}$ values for the two different models). We note also that due to the complex geometry the quoted energy values do not necessarily represent the true minima of the interaction potential---see \SIapp{4}.

Below we present simulations of a system containing $N\!=\!1000$~model HP1s and an $L\!=\!1000$ bead polymer. For simplicity, we consider a homogeneous polymer where all beads can bind HP1, i.e., it represents a section of H3K9me3 modified chromatin.
For the multivalent HP1 model we ran long simulations to obtain equilibrium configurations, as detailed in \SI{}; in several cases we ran additional test simulations to check that the configurations are indeed representative of equilibrium, \SIapp{5}. For the limited valence HP1s the system displayed long-lived non-equilibrium metastable states (see below).
We confine all components of the system in a cubic box of size $l_x = 35\sigma$ (approximately equal to the radius of gyration of the polymer as predicted by the worm-like chain model) by including a ``wall potential''.
While the confinement reduces the entropy of the system by forbidding some extended polymer configurations, it also prevents the polymer from interacting with its periodic image. 

\subsection*{Model 1: Multivalent protein-protein interactions}

With this version of the model, when the HP1-chromatin interaction energy $\epsilon_{\rm HC}$ is small the proteins behave like a standard phase separating system [Model B~\cite{ChaikinLubensky,cates_tjhung_2018}]. When $\epsilon_{\rm HC}$ is larger there is more interesting behaviour. We summarise the emerging regimes in the simulation snapshots and phase diagram in Figs.~\ref{fig:iso_phase_diag}(a-c). 

  When $\epsilon_{\rm HC}<12k_BT$, a phase transition between a uniform mixed phase and a separated phase takes place as $\epsilon_{\rm HH}$ increases. Above a critical $\epsilon_{\rm HH}$ value a roughly spherical cluster, or ``droplet'', of HP1 forms. 
We call this the \textit{dense droplet} regime. By measuring the density $\rho$ of HP1s inside and outside of the droplet, we can also map out the phase diagram on the $\rho$-$\epsilon_{\rm HH}$ plane [Fig.~\ref{fig:iso_phase_diag}(c) top, see \SIapp{6} for details].

\begin{figure*}[ht]
\includegraphics{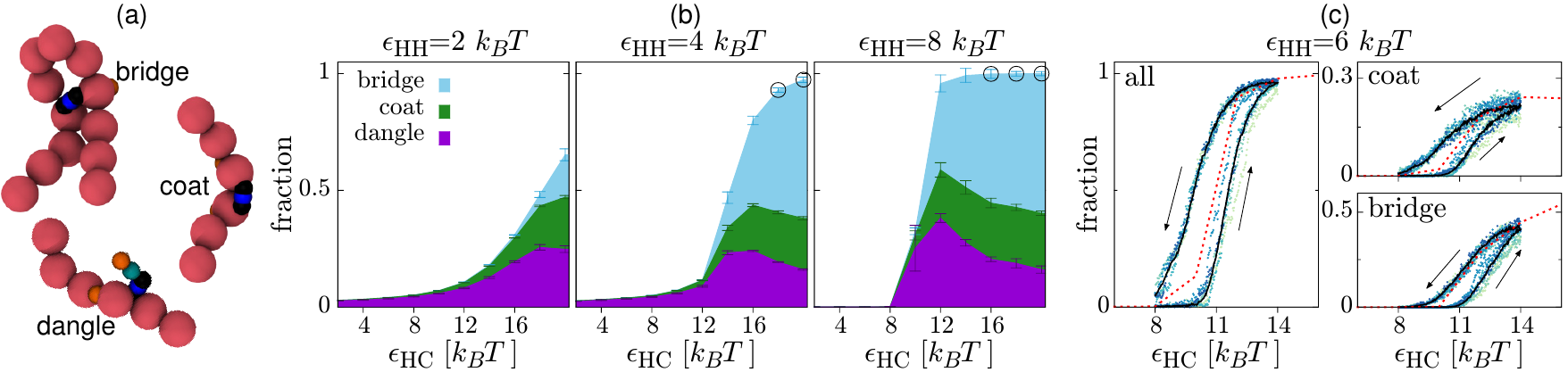}
\caption{\textbf{Protein-chromatin binding modes.} (a) Simulation snapshots of a single model HP1 and a short section of chromatin. The three different binding modes are depicted. (b) Plots showing the fraction of the $N\!=\!1000$ proteins bound to the chromatin in each mode for different interaction energies. The height of each coloured region indicates the proportion of proteins, with the regions stacked on top of each other. In this way the height of the total coloured region indicates the fraction of proteins bound in any mode $f_{\rm tot}$. Values are obtained from averaging over 4 simulations of duration $5\!\times\!10^3\tau$, and error bars show the standard error in the mean. Black circles around points indicate where bridging is the dominant binding mode (more than half of the bound proteins). (c) Plots showing the fraction of proteins bound (total or in the indicated mode) during a simulation where the HP1-chromatin interaction energy was slowly varied: starting at $\epsilon_{\rm HC}=8k_BT$ it was increased to $\epsilon_{\rm HC}=14k_BT$ over $2\!\times\!10^4\tau$, before being decreased again over the same time interval. The HP1-HP1 interaction energy was kept fixed at $\epsilon_{\rm HH}=6k_BT$. Blue-green points show data from 12 independent simulations, each in a different colour. Black solid lines show an average over these simulations. The red dotted line shows values obtained from equilibrium simulations as in (b).}
\label{fig:iso_bindingmode}
\end{figure*}

For small HP1-HP1 interaction energies, $\epsilon_{\rm HH} < 5k_BT$, there is no droplet. Increasing the HP1-chromatin attraction leads to HP1s becoming bound to the polymer, and there is a smooth increase of the fraction bound as $\epsilon_{\rm HC}$ increases. For large $\epsilon_{\rm HC}$ there are sufficient HP1s bound such that the region occupied by the chromatin has a higher than average protein density, and the surroundings have a lower than average protein density [green region in Figs.~\ref{fig:iso_phase_diag}(a-b)]. In this sense there is a phase separation, however this regime is profoundly different from the dense droplet phase: a significant fraction of the proteins remain unbound, while the rest tend to ``coat'' the polymer. Hence, we refer to it as the \textit{coating} regime.

When both $\epsilon_{\rm HH}$ and $\epsilon_{\rm HC}$ are large [blue region in Figs.~\ref{fig:iso_phase_diag}(a-b)] a protein droplet forms, but now the polymer is also absorbed into it. Or in other words, the droplet compacts the polymer. We call this the \textit{absorbing droplet} regime. Interestingly, the polymer is absorbed to a different degree depending on the precise values of the interaction energies [compare snapshots at $\epsilon_{\rm HC}=20k_BT$ and different $\epsilon_{\rm HH}$ in Fig.~\ref{fig:iso_phase_diag}(a), where different amounts of chromatin extend out from the droplet]. As before, measurements of HP1 density inside and outside of the droplet allows us to construct the $\rho$-$\epsilon_{\rm HH}$ phase diagram for large $\epsilon_{\rm HC}$, on which we can also identify the coating regime [Fig.~\ref{fig:iso_phase_diag}(c) bottom, and see \SIapp{6}]. There is a further new region on this phase diagram (the shaded stripe) where a droplet forms only due to HP1-chromatin interactions (i.e., $\epsilon_{\rm HH}$ is not large enough for a protein droplet to form on its own). We discuss this in more detail below.

To characterise these regimes more quantitatively [and to determine the the positions of the lines in Fig.~\ref{fig:iso_phase_diag}(b)], we measure the local protein density by splitting the simulation box into $N_{\rm sb}$ sub-boxes of volume $V_{\rm sb}$. If there are $N_i$ HP1s in the $i$th sub-box, the local density is ${\rho_i = N_i/V_{\rm sb}}$.
To quantify the level of phase separation we then consider a `separation depth' parameter~\cite{vladimirova1998diffusion} defined as
\begin{equation}
\phi_{\rm sep} =\frac{1}{N_{\rm sb}} \sum_{i=1}^{N_{\rm sb}} \frac{\rho_i - \rho}{\rho^* - \rho},
\label{sepdepth}
\end{equation}
where $\rho = N/l_x^3 $ is the overall number density of HP1s, and $\rho^*$ is a reference density which takes the value $\rho_+$ when $\rho_i > \rho_+/2$ and $\rho_- = 0$ otherwise. We use $\rho_+=0.5$ and $N_{\rm sb}=125$, chosen so as to be optimal for distinguishing the different regimes, and leading to $\phi_{sep} \rightarrow 1$ on droplet formation. 
Figure~\ref{fig:iso_phase_diag}(d) shows how $\phi_{\rm sep}$ varies with $\epsilon_{\rm HH}$, for different values of $\epsilon_{\rm HC}$. For $\epsilon_{\rm HC} \leq 12k_BT$ the points sit on top of each other [Fig.~\ref{fig:iso_phase_diag}(d) inset], and we observe a sharp crossover (at $\epsilon_{\rm HH}\approx4.5k_BT$) from $\phi_{\rm sep} \sim 0.15$ in the mixed phase to $\phi_{\rm sep} \sim 1$ in the dense droplet phase. As noted above, for these values of the energy the model behaves qualitatively the same as, e.g., interacting Brownian colloids~\cite{Zaccarelli2007}, and we expect a first-order phase transition in the thermodynamic limit (Model B). 
We use a value of $\phi_{\rm sep}=0.5$ to set the position of the orange line in Fig.~\ref{fig:iso_phase_diag}(b). 
As $\epsilon_{\rm HC}$ increases, this line shifts to the left---we discuss this interesting regime further below. 
 Figure~\ref{fig:iso_phase_diag}(e) shows that for small $\epsilon_{\rm HH}$ the separation depth is independent of $\epsilon_{\rm HC}$ throughout the uniform phase ($\phi_{\rm sep} \sim 0.15$), before increasing at larger $\epsilon_{\rm HC}$ in the coating or absorbing droplet regimes; we take the value of $\epsilon_{\rm HC}$ at which $\phi_{\rm sep}$ starts to increase as the point where the system enters the coating regime [green dashed line in Fig.~\ref{fig:iso_phase_diag}(b)].
For $\epsilon_{\rm HH}> 6k_BT$ the separation depth $\phi_{\rm sep}\sim1$, independently of $\epsilon_{\rm HC}$; i.e., this parameter cannot differentiate between droplets and absorbing droplets.
  
We now consider the nature of the interactions between the HP1 dimers and the chromatin. 
Since each model HP1 dimer can interact with the polymer via two distinct domains (the CDs), they can bind in three different modes [Fig.~\ref{fig:iso_bindingmode}(a)]. First, an HP1 could bind through only one of the CDs; we call this ``dangling'', since it leaves one free CD. Second, the CDs could both bind to the chromatin at adjacent ($|i-j| < 2$) polymer beads; we call this ``coating''. Finally, if the CDs interact with polymer beads which are separated along the chain ($|i-j| \geq 2$), then the protein is ``bridging''.
As detailed in Ref.~\cite{Brackley2020JPCM}, the shape of the protein determines its likelihood to bind in each mode: bridging incurs an entropic penalty (due to polymer looping), so unless the shape of the protein specifically disfavours coating, the coating mode is favourable. This is the case here: in the absence of protein-protein interactions we mainly observe coating. The ability of real HP1 dimers to form bridges between distant chromatin regions remains unclear; however, cryo-electron microscopy~\cite{Machida2018} and detailed molecular simulations~\cite{Watanabe2018} have indicated that HP1 can readily sit between adjacent nucleosomes, suggesting that (at least under dilute conditions) coating may well dominate. 

In Fig.~\ref{fig:iso_bindingmode}(b) we plot the fraction of bridging, coating and dangling proteins as a function of $\epsilon_{\rm HC}$. If we consider the total fraction of proteins bound to the polymer $f_{\rm tot}$, at small $\epsilon_{\rm HH}\!=\!2k_BT$, $f_{\rm tot}$ increases smoothly with $\epsilon_{\rm HC}$. 
Coating and dangling are the dominant binding modes; the BIA is therefore not in effect, and we do not observe BIPS or chromatin compaction. At large $\epsilon_{\rm HH}$, where there is a droplet, $f_{\rm tot}$ increases very sharply as $\epsilon_{\rm HC}$ is increased and the polymer becomes absorbed into the droplet [the curve becomes steeper from left to right in the panels of Fig.~\ref{fig:iso_bindingmode}(b)]. This could indicate the presence of a first-order phase transition in the thermodynamic limit. Within the absorbing droplet regime we also observe that the fraction of bridging proteins increases with $\epsilon_{\rm HC}$, and it becomes the dominant mode of binding when both interactions are strong. The main driver of this is that as $\epsilon_{\rm HC}$ increases, more of the polymer becomes absorbed, and so the likelihood of two distant regions being close enough together for bridges to form increases. 
We also performed some simulations where the HP1-chromatin interaction was increased slowly (after starting in an equilibrium configuration), before being decreased again [Fig.~\ref{fig:iso_bindingmode}(c)]. As detailed further in \SI{}, the system displays hysteresis as the polymer becomes absorbed and then re-emerges from the droplet.  
These observations suggest that in the limit of a large droplet the system would show a first-order transition as $\epsilon_{\rm HC}$ increases, to a phase where the polymer is fully absorbed; in our small system we instead observe an extended co-existence regime where the polymer is only partially absorbed. We use the point where $f_{\rm tot}=0.5$ to set the position of the solid green line in Fig.~\ref{fig:iso_phase_diag}(b).

\begin{figure}[!t]
\includegraphics{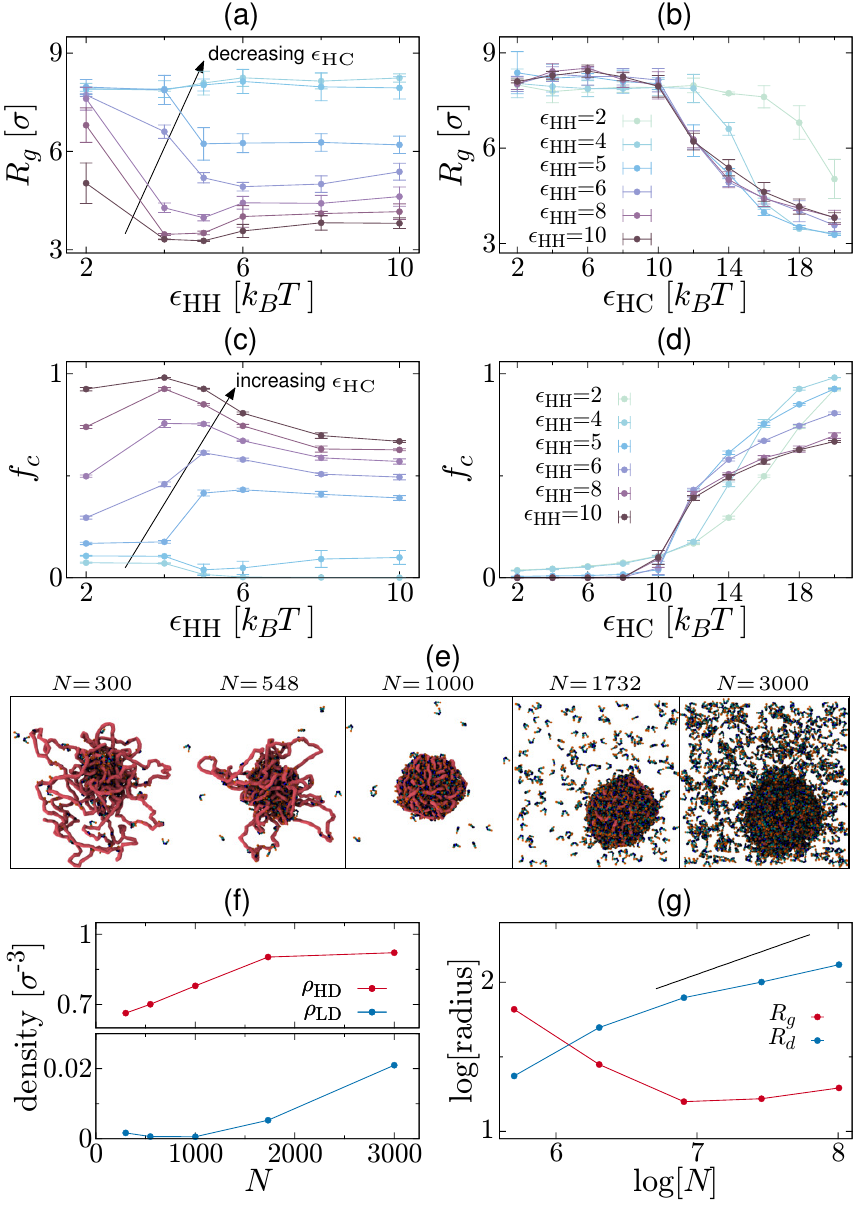}
\caption{\textbf{HP1-chromatin interactions, chromatin compaction, and droplet dynamics.} (a-b) Plots showing how the radius of gyration of the polymer representing the chromatin segment depends on the interaction energies. In (a) from top to bottom curves are for $\epsilon_{\rm HC}$ values between 8 and 20$k_BT$ increasing in steps of 2$k_BT$ from top to bottom. Points are obtained from an average of 4 independent simulations; error bars show the standard error in the mean, and connecting lines are a guide to the eye.
(c-d)~Plots showing how the fraction of chromatin beads which are bound by proteins $f_{c}$, depends on the interaction energies. 
In (c) from bottom to top, curves are for $\epsilon_{\rm HC}$ values between 8 and 20$k_BT$, increasing in steps of $2k_BT$. 
(e) Snapshots from simulations with $\epsilon_{\rm HH}=4k_BT$ and $\epsilon_{\rm HC}=20k_BT$ but with different numbers of proteins $N$ as indicated. (f) Plot showing how the protein densities within the high and low density phases (inside and outside the droplet) vary with the number of proteins. (g) Plot showing how the radius of gyration of the polymer $R_g$ and radius of the droplet $R_d$ vary with $N$, shown in log-log scale. The black line has a slope $1/3$, which is how the droplet radius would scale in a standard phase separating system. 
\label{fig:iso_polmer}}
\end{figure}
 
One proposed function of HP1 \textit{in vivo} is to compact heterochromatin. The ability of our model proteins to compact the chromatin can be probed by measuring its radius of gyration, defined as 
\begin{equation}
R_g^2 =\frac{1}{L} \sum_{i=1}^{L} (\mathbf{r}_i - \bar{\mathbf{r}})^2,
\label{rg}
\end{equation}
where $\mathbf{r}_i$ is the position of the $i$th chromatin bead, and $\bar{\mathbf{r}} = (1/L)\sum_i \mathbf{r}_i$. Figures~\ref{fig:iso_polmer}(a-b) show how $R_g$ depends on the interaction energies. Interestingly, $R_g$ can vary non-monotonically as $\epsilon_{\rm HH}$ increases; similar behaviour is observed in the fraction of polymer beads bound by proteins, $f_c$ [Fig.~\ref{fig:iso_polmer}(c,d)]. The reason for this non-monotonicity is strikingly apparent in the top row of snapshots in Fig.~\ref{fig:iso_phase_diag}(a): in the leftmost snapshot the polymer is swollen, in the second from the left it is fully absorbed into a protein droplet (small $R_g$ and large $f_c$), but in the two right-hand snapshots the polymer is only partially absorbed into the droplet ($R_g$ increases again, while $f_c$ decreases). That the amount of absorbed polymer varies so widely within the absorbing droplet regime is likely due to competition between different contributions to the free energy. While HP1-chromatin binding represents a reduction in free energy, this is offset by the reduction in entropy due to the compaction/confinement of the polymer within the droplet. Increasing $\epsilon_{\rm HC}$ increases the amount of chromatin absorbed as the entropic loss is overcome. On the other hand, the presence of the polymer within a droplet will reduce the number of HP1-HP1 interactions due to steric effects; so increasing $\epsilon_{\rm HH}$ \textit{decreases} the amount of chromatin absorbed (effectively the polymer is `squeezed out' of the droplet).

Finally in this section, we consider intermediate values of the HP1-HP1 interaction strength, $\epsilon_{\rm HH}\approx4k_BT$, where we observe the most interesting behaviour. Here, in the absence of chromatin interactions there is no droplet formation and $\phi_{\rm sep}$ is small. However, we note that as $\epsilon_{\rm HC}$ increases, the orange line in Fig.~\ref{fig:iso_phase_diag}(b) moves to the left, so droplets \textit{can} form at $\epsilon_{\rm HH}\approx4k_BT$ \textit{if} the protein-chromatin interaction energy is large enough. In other words, HP1-chromatin attraction promotes protein aggregation. This can be understood as follows: when $\epsilon_{\rm HC}$ is large enough, a significant number of HP1s become localised to the polymer and these tend to bind in the coating mode. Then, intermediate HP1-HP1 interactions are sufficient to allow extended {chromatin-HP1-HP1-chromatin} bridges to form. The BIA is therefore in effect, leading to chromatin compaction and protein clustering; we note that this is the \textit{only} region of the phase diagram where the BIA is really in effect and a true BIPS is observed. When both $\epsilon_{\rm HH}$ and $\epsilon_{\rm HC}$ have intermediate values we observe an absorbing (BIPS) protein droplet \textit{and} coating of the chromatin which emerges from the droplet [cross-hatch shaded region in Fig.~\ref{fig:iso_phase_diag}(b)]. 

\subsection*{Varying protein density}

We now consider the effect of the overall protein density for the multivalent HP1s. As expected, for large $\epsilon_{\rm HH}$, we observed the same behaviour as a standard (Model-B) phase separation: increasing overall protein density leads to an increase in the size of the droplet, while the density of proteins within it remains constant. If $\epsilon_{\rm HC}$ is also large, the amount of absorbed chromatin grows with the size of the droplet (\SIFig{8} and see \SIapp{9}). Interestingly, once the droplet is large enough to fully absorb the polymer, further increasing $N$ (and further increase of the droplet size) does not lead to a swelling of the polymer: the ratio $R_g/R_d$ continues to decrease with $N$ [\SIFig{8}(e)].

A strikingly different behaviour is observed for intermediate $\epsilon_{\rm HH}$ (the region where where the BIA is in effect, i.e, where a droplet only forms due to the presence of the polymer). Figure~\ref{fig:iso_polmer}(e) shows snapshots for $\epsilon_{\rm HH}=4k_BT$ and $\epsilon_{\rm HC}=20k_BT$ with different numbers of proteins. It is immediately clear that the density of proteins within the two phases varies with $N$ [also Fig.~\ref{fig:iso_polmer}(f)]. This can be rationalised as follows. For small $N$ a protein droplet forms on the polymer via the BIA. This droplet is rather `loose', and as $N$ increases, more space within the droplet becomes filled with proteins and the density ($\rho_{\rm HD}$) increases. At the same time more polymer becomes absorbed and the droplet grows [$R_g$ decreases, and the droplet diameter $R_d$ increases, Fig.~\ref{fig:iso_polmer}(g) and \SIFig{9}]. When $N\approx1000$ all of the polymer is absorbed, and $R_g$ reaches a minimum; as $N$ and $R_d$ increase further the polymer can swell slightly. At some point the droplet density plateaus, and adding further proteins instead leads to an increase in the density of proteins outside the droplet ($\rho_{\rm LD}$). The droplet still grows with $N$, but more slowly than in a standard phase separation (where $R\sim N^{1/3}$); the fraction of proteins binding the polymer in the bridging mode decreases at the expense of the other two modes (\SIFig{9} and see \SIapp{9}).

In summary, for the narrow range of parameters where phase separation only occurs in the presence of the polymer, we find the surprising result that the density of the phases ($\rho_{\rm HD}$ and $\rho_{\rm LD}$) depends on the overall protein density [shaded band in Fig.~\ref{fig:iso_phase_diag}(d)]. This has important implications for protein-chromatin interaction \textit{in vivo} (see Discussion).

\subsection*{Model 2: Limited valence protein-protein interactions}

\begin{figure}[t]
\includegraphics{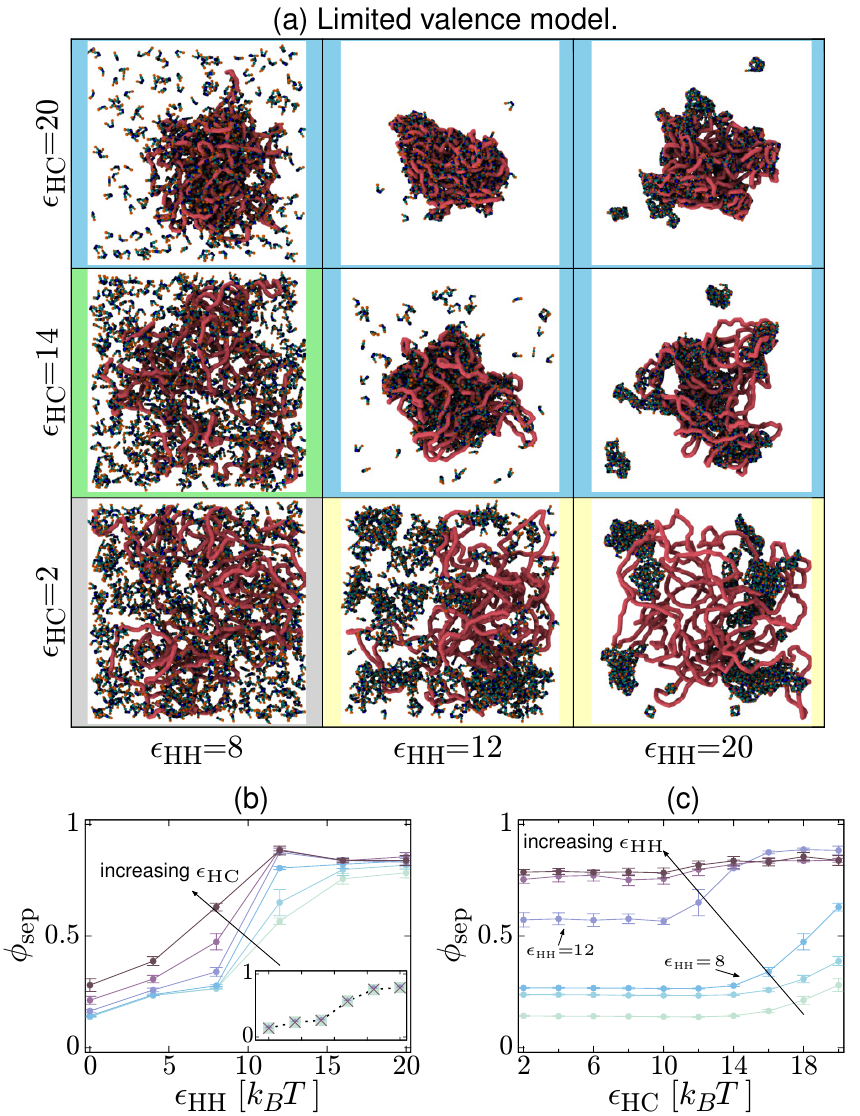}
\caption{\textbf{The limited valence HP1s display similar behaviour regimes.} (a) Snapshots are shown for simulations of the the limited valence HP1 model with different HP1-HP1 and HP1-chromatin interaction energies. Border colours indicate similarity to the different regimes observed for the multivalent model in Fig.~\ref{fig:iso_phase_diag} (see also \SIFig{12}).
(b) Plot showing how the separation depth parameter varies with $\epsilon_{\rm HH}$ for different values of $\epsilon_{\rm HC}$ for the limited valence model. Data for $\epsilon_{\rm HC}$ between 10 and $20k_BT$ increasing in steps of 2$k_BT$ are shown in the main plot. The inset shows that points for $\epsilon_{\rm HC}=6,8$, and $10k_BT$ sit on top of each other. (c)~Similar plot showing $\phi_{\rm sep}$ as a function of $\epsilon_{\rm HC}$. Curves are for different values of $\epsilon_{\rm HH}$ between 0 and 20$k_BT$ increasing from bottom to top in steps of $4k_BT$. \label{fig:iso_polmerLV} }
\end{figure}

In this model the HP1 dimer-dimer interactions have a limited valence, i.e., exactly one hinge domain can interact with exactly one NTE domain at a time. These HP1s behave like classic patchy particles, which have been studied extensively using both simulations~\cite{Teixeira2017,Bianchi2008} and experiments~\cite{Fusco2013,Zhang2004}. Patchy particles have a rich phase diagram which includes a low-density equilibrium gel phase, and ``closed loop'' structures (where a set of particles form a structure where all patches are bound). Technically these different but co-existing equilibrium states are only present at zero temperature, but the structures can also exist as very long lived non-equilibrium metastable states for non-zero temperatures~\cite{Zaccarelli2007}. In our simulations a gel phase is precluded since the system is confined. In the previous section we specifically considered equilibrium configurations; here we study the metastable states obtained when the system is quenched by instantaneously switching on both protein-protein and protein-chromatin interactions. Specifically, we start from an equilibrium configuration for $\epsilon_{\rm HH},\epsilon_{\rm HC}=0$, switch on interactions and run for $10^4\tau$ (where $\tau$ is the simulation time unit); after this time the measured quantities ($f_c$, $\phi_{\rm sep}$, etc.) have stopped systematically varying. Steady state values of these quantities are then obtained by averaging over a further $10^4\tau$ simulation. 

Typical snapshots are shown in Fig.~\ref{fig:iso_polmerLV}(a). Similar behaviour is observed as for the multivalent interaction model. At low $\epsilon_{\rm HH}$ we have the same mixed and coating regimes. For low $\epsilon_{\rm HC}$, as $\epsilon_{\rm HH}$ increases, we go from the mixed phase to an aggregate phase. Unlike the multivalent model, here the aggregates are not spherical; instead multiple irregularly shaped clusters form. We also see small closed clusters of HP1s where all hinge and NTE domains are bonded  When both $\epsilon_{\rm HC}$ and $\epsilon_{\rm HH}$ are large, many of the aggregates become associated with the polymer, which becomes compacted. Some smaller clusters remain detached from the polymer. Measurements of clusters and sub-clusters (see \SI{}) show that these have a fractal dimension less than 3, as would be expected in a gel.  It is important to reiterate that these are long-lived metastable, dynamically arrested structures, and do not represent a true equilibrium of the system. Using a different quenching or annealing procedure with the same parameters leads to different relative abundances of the different types of aggregate (see \SIapp{11}).

As before we measure the separation depth $\phi_{\rm sep}$ as a function of the two interaction energies. Figure~\ref{fig:iso_polmerLV}(b) shows that the behaviour is again similar to the multivalent model in that $\phi_{\rm sep}$ increases with $\epsilon_{\rm HH}$. However, the largest $\phi_{\rm sep}$ values are smaller than in the multivalent case, consistent with several protein aggregates of different size forming, rather than a single phase separated droplet. 
There is also a regime where proteins aggregate only when the interaction with the chromatin is strong enough, though it is less clear than for the multivalent model. Specifically, for $\epsilon_{\rm HH}=8k_BT$ there is a cluster only when $\epsilon_{\rm HC}$ is large, but $\phi_{\rm sep}$ only reaches intermediate values [Figs.~\ref{fig:iso_polmerLV}(a) and (c)]. For $\epsilon_{\rm HH}=12k_BT$, $\phi_{\rm sep}$ has an intermediate value just less than 0.6 for a broad range of $\epsilon_{\rm HC}$ values [Figs.~\ref{fig:iso_polmerLV}(c)], behaviour which is not observed in the multivalent model. This arises because while clusters do form, there are many of them; they are also highly dynamic, continually forming, dissolving, merging and breaking apart~\cite{Zaccarelli2007}.

 We again measure the fraction of proteins bound in different modes [\SIFig{13}], the fraction of polymer beads bound by proteins [\SIFig{14}(a-b)], and the radius of gyration of the polymer [\SIFig{14}(c-d)] as a function of the two interaction energies. The behaviour is broadly similar to the multivalent model, but the limited valence proteins are less able to compact the polymer, and the chromatin ``looping out'' (which was observed in the multivalent case when both energies were large) does not tend to occur here. Instead, most of the polymer is associated with the irregularly shaped cluster. 

\begin{figure}[!t]
\includegraphics{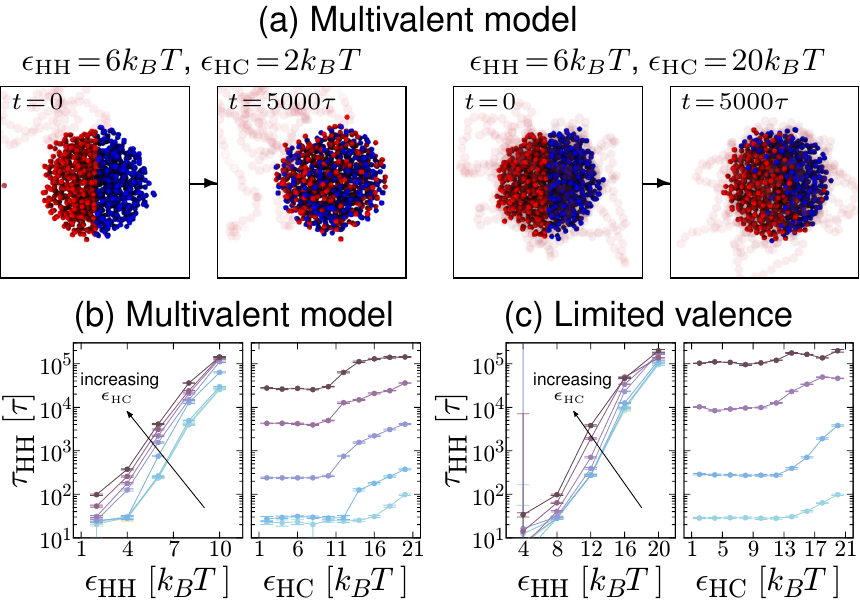}
\caption{\textbf{Dynamics of protein-protein interactions.} (a) Simulation snapshots showing protein mixing over time for two different parameter sets for the multivalent HP1 model. At $t=0$ we colour each protein according to which half of a droplet it is in; for simplicity only one bead from each protein is shown, and the polymer is shown transparent. Images at $t=0$ and $t=5\times10^3\tau$ are shown. For $\epsilon_{\rm HH}=6k_BT$, $\epsilon_{\rm HC}=2k_BT$ (left) the system is in the droplet phase and the proteins are not associated with the chromatin, while $\epsilon_{\rm HH}=6k_BT$, $\epsilon_{\rm HC}=20k_BT$ (right) is in the absorbing droplet regime. (b) Plots for the multivalent model showing the HP1 bond decorrelation time constant $\tau_{\rm HH}$ in simulation time units $\tau$ as a function of the HP1-HP1 and HP1-chromatin interaction energies. Points show $\tau_{\rm HH}$ values obtained from fits as detailed in \SIapp{12} and \SIFig{15}. Error bars show the error obtained from the fit and connecting lines are a guide for the eye. In the left plot from bottom to top (light to dark colours) lines are for increasing $\epsilon_{\rm HC}$ (between 2 and 20$k_BT$ in steps of 2$k_BT$), while in the right plot bottom to top shows increasing $\epsilon_{\rm HH}$ (between 2 and 10$k_BT$ in steps of 2$k_BT$). (c) Similar plots are shown for simulations with the limited valence model. In the left plot lines are for $\epsilon_{\rm HC}$ values between between 2 and 20$k_BT$ in steps of 2$k_BT$; in the right plot lines are for $\epsilon_{\rm HH}$ between 8 and 20$k_BT$ in steps of 4$k_BT$. Points for $\epsilon_{\rm HH}=4k_BT$ are not shown since the data were noisy and the fitted $\tau_{\rm HH}$ values had very large errors (see \SI{}). \label{fig:dynamics}}
\end{figure}

\subsection*{Protein dynamics}

So far we have considered structural properties of the protein clusters for each of the two models. Here we consider protein dynamics. This is often studied \textit{in vivo} using fluorescence recovery after photo-bleaching (FRAP) experiments: the time scale of fluorescence recovery of a protein droplet gives a measure of how quickly proteins are exchanged between the droplet and the soluble (unbleached) pool. The internal dynamics of a droplet can also be probed by photo-bleaching half of the droplet: tracking fluorescence in the bleached and un-bleached halves gives information on the relative time scales of mixing within the droplet and exchange with the soluble pool~\cite{Erdel2020}. A similar effect can be observed visually in simulations by colouring proteins according to which half of a droplet they are in, and then watching how the colours mix in time. Two examples for the multivalent proteins (Model 1) with different $\epsilon_{\rm HC}$ values are shown in Fig.~\ref{fig:dynamics}(a).  
We observe that for small $\epsilon_{\rm HC}=2k_BT$ (where the droplet is not associated with chromatin) there is a high degree of mixing over the duration of the simulation. Interestingly, for larger $\epsilon_{\rm HC}=20k_BT$ (but the same value of $\epsilon_{\rm HH}=6k_BT$) where the chromatin is absorbed into the droplet, we find that the colours mix to a much lesser extent.

More quantitatively, we can measure how the proteins change their binding partners during a given time interval $\Delta t$ by defining a bond-bond correlation function
\begin{equation}\label{dynpar}
\nu_{\rm HH}(\Delta t)= \frac{\langle y_{ij}(t+\Delta t) \; y_{ij}(t)\rangle - \langle y_{ij}(t)\rangle^2 }{\langle y_{ij}(t)^2\rangle - \langle y_{ij}(t)\rangle^2},
\end{equation}
where $y_{ij}(t)$ has a value of $1$ if proteins $i$ and $j$ are interacting at time $t$, and $0$ otherwise (proteins are said to interact if a hinge or NTE domain on protein $i$ is within the interaction range from a NTE or hinge on protein $j$). Angle brackets denote an average over time $t$, repeat simulations, and all possible $i,j$ pairs of proteins. As detailed in \SI{}, the shapes of the $\nu_{\rm HH}(\Delta t)$ curves suggest that there are multiple time scales involved in the decorrelation; it is nevertheless possible to extract a single overall decorrelation time, $\tau_{\rm HH}$ (the typical time for all proteins to change their binding partners, see \SIapp{12} and \SIFig{15} for details). In Figs.~\ref{fig:dynamics}(b-c) we plot $\tau_{\rm HH}$ as a function of the different interaction energies, for the multivalent and limited valence models respectively. For the multivalent model [Fig.~\ref{fig:dynamics}(b)], $\tau_{\rm HH}$ grows with the HP1-HP1 interaction energy roughly exponentially (roughly linear on the log-linear plot). The exception is where both $\epsilon_{\rm HH}$ and  $\epsilon_{\rm HC}$ are small, where there is no droplet (proteins only interact transiently) and $\nu_{\rm HH}$ drops to zero almost instantaneously. More interestingly, as $\epsilon_{\rm HC}$ is increased there is a clear step change in $\tau_{\rm HH}$ where the system goes from the droplet to the absorbing droplet regime. In other words, consistent with the snapshots in Fig.~\ref{fig:dynamics}(a), the presence of the chromatin within the droplet leads to a dramatic slow-down of protein dynamics ($\tau_{\rm HH}$ increases by almost an order of magnitude). As $\epsilon_{\rm HC}$ is further increased there again a roughly exponential dependence of $\tau_{\rm HH}$ on $\epsilon_{\rm HC}$ [linear increases for $\epsilon_{\rm HC}\gtrapprox12k_BT$ on the log-linear scale in the left panel of \ref{fig:dynamics}(b)].  
While mapping between simulation and physical time units is not straightforward (see \SIapp{3}), the time interval between the images in Fig.~\ref{fig:dynamics}(a) is of the order 3-4 minutes [compare to the 30-50s time scale measured experimentally for HP1 recovery after photobleaching in Ref.~\cite{Muller2009}, measuring exchange with the soluble pool].

The limited valence proteins [Fig.~\ref{fig:dynamics}(c)] show similar behaviour. The decorrelation time $\tau_{\rm HH}$ again grows roughly exponentially with $\epsilon_{\rm HH}$ (though there is some deviation from this for large $\epsilon_{\rm HC}$). The right-hand panel in Fig.~\ref{fig:dynamics}(c) shows that again  $\tau_{\rm HH}$ starts to increase with $\epsilon_{\rm HC}$ when the proteins start to interact with the polymer, at least for intermediate values of $\epsilon_{\rm HH}$. There is not such a pronounced step-change as in the multivalent case, likely due to the presence of protein clusters which do not interact with the polymer. For larger $\epsilon_{\rm HH}$, interaction with the polymer has a much smaller effect. 

\section*{Discussion}

In this paper we have studied the behaviour of simple model proteins interacting with a bead-and-spring polymer model for chromatin. We considered rigid bodies composed of spheres that represent different protein domains which interact attractively with each other or with chromatin. The domain structure was based on that of HP1, but our goal was to obtain insight on the interplay between protein-protein and protein-chromatin interactions in general. 

HP1 has been shown to undergo liquid-liquid phase separation (LLPS) \textit{in vitro}~\cite{Larson2017,Strom2017}. That result lead to the suggestion that LLPS could play a major role in formation of chromatin associated protein foci \textit{in vivo}. Here, we also considered that many protein complexes bind chromatin multivalently, and so the bridging-induced attraction (BIA) can also play a role.
 Using a model with multivalent protein-protein interactions, we found that in the absence of protein-chromatin interactions, increasing the protein-protein interaction strength $\epsilon_{\rm HH}$ led to liquid droplet formation (Model B). Increasing protein-chromatin attractive interactions lead to a sharp crossover to a regime where the chromatin is absorbed into the droplet (with indications that there is a first-order phase transition in the thermodynamic limit). Importantly, the level of chromatin absorption depended on both interaction energies, and the number of proteins/size of the droplet. For most of the parameters studied, a significant fraction of the chromatin ``looped out'' from the droplet [the looping statistics of a similar situation have been studied in Ref.~\cite{Broedersz2014}].  This suggests that precise parameter tuning would be required for protein-protein attraction (LLPS) alone to mediate chromatin associated protein droplet formation and chromatin compaction/isolation \textit{in vivo}. 
We would expect that chromatin regions with the correct histone modifications for protein binding act as droplet nucleation points. It would be interesting in the future to study the case where there were multiple such nucleation points (which are, e.g, kept spatially separated due to nuclear structure), or where these change in time (e.g., if histone modifications change dynamically as genes are activated).

An interesting regime in our multivalent protein simulations is for intermediate values of $\epsilon_{\rm HH}$, where a droplet only forms if $\epsilon_{\rm HC}$ is large enough. In other words, phase separation is promoted by interaction with chromatin; this can be viewed as {chromatin-HP1-HP1-chromatin} bridges enabling the BIA to drive protein clustering. In this regime we also see a dependence on the overall protein density $\rho$ which is fundamentally different to standard Model B phase separation. The density of proteins within and outside the droplet depends on $\rho$, and the droplet volume grows sub-linearly as $\rho$ increases. This behaviour originates from the formation of a ``loose'' protein cluster on the chromatin for small $\rho$, which can ``fill up'' as proteins are added to the system; at larger $\rho$, sites on the chromatin become saturated, so as more proteins are added these instead remain unbound (increasing the density in the protein poor region). This is reminiscent of recent work showing that varying the overall concentration of the nucleophosmin protein (a key component of nucleoli, which form via LLPS) leads to variation in its density both inside and outside the nucleolus; in that system there are multiple phase-separating components which leads to a complicated high-dimensional phase diagram~\cite{Riback2020}. 

Whether HP1 undergoes LLPS \textit{in vivo} is still a topic of debate, and there are conflicting observations~\cite{Erdel2020,Keenen2021}. One recent study showed that over-expression of HP1 in mouse \textit{does not} lead to an increase in the size of foci, but instead the protein density within the foci increases~\cite{Erdel2020}. While inconsistent with a classic phase separation mechanism, this observation is compatible with the intermediate $\epsilon_{\rm HH}$ (BIA) regime discussed above. The same work used FRAP experiments to show that mixing within HP1 droplets is slower than exchange with the soluble pool~\cite{Erdel2020}. 
Our simulations showed nearly an order of magnitude slow-down in protein dynamics when the chromatin is absorbed into the droplet. Due to the small system size, the rate of exchange between the droplet and pool is difficult to measure in our simulations; however, the time scales for exchange with the pool and internal mixing are likely similar. 
The concurrent slow internal mixing and fast exchange with the pool [of the order 10~s~\cite{Muller2009}]  is therefore not reproduced in the simulations (though see below). Nevertheless, our results suggest that care should be taken when interpreting FRAP measurements of internal mixing: slow mixing may be due to the presence of chromatin, and does not necessarily preclude LLPS.
Ref.~\cite{Erdel2020} also showed that removal of the H3K9me3 histone modification leads to loss of HP1 co-localisation, but the heterochromatin foci remain intact (inconsistent with HP1 begin a driver of heterochromatin body formation). Other work~\cite{Feng2020} has suggested that while HP1 may not be necessary to compact large satellite repeat heterochromatin regions, it \textit{is} required to compact and silence smaller H3K9me3 marked segments within otherwise active regions. The function of HP1 clearly still not well understood. Nevertheless, the simplicity of our model proteins means that our results are also likely to be relevant for other proteins or complexes. An example is the H1 linker histone, associated with chromatin compaction and gene repression, which has been shown to phase separate in the presence of DNA~\cite{Shakya2020}, and has multiple DNA binding sites as well as interacting with the core histones~\cite{White2016}. 

The limited valence model showed similar regimes to the multivalent case, but instead of a spherical droplet the proteins formed fractal clusters [similar to the structures formed by patchy particles~\cite{Teixeira2017,Bianchi2008}]. The limited valence HP1s could also form a gel in simulations with a higher density and periodic boundaries (\SIFig{11}). 
NMR spectroscopy experiments have shown that phosphorylated HP1$\alpha$ forms a gel \textit{in vitro} if condensates are left for around seven days~\cite{Ackermann2019}. 
A possible explanation for this is that first, weak multivalent interactions between the disordered domains drive LLPS~\cite{Martin2020}, with gelation occurring on longer time scales as these rearrange or fold~\cite{Lin2018}. This is consistent with our result that small changes to the nature of the protein-protein interactions can lead to large morphological differences in  the resulting condensates (droplets \textit{vs.} fractal clusters), and could be related to the observation that HP1 foci can have quite different properties in different cell types~\cite{Dialynas2007}.
There are also broader implications: LLPS and gelation have been associated with amyloid formation in neurodegenarative disease~\cite{Pytowski12050}, and it is possible that formation of gels or fractal clusters of chromatin associated proteins may also be pathological. For example, small disease associated mutations in the protein MeCP2 (also associated with heterochromatin) were recently shown to prevent it from undergoing LLPS~\cite{Li2020}.

Finally, we note that our model proteins are ``poor bridgers'', which tend to ``coat'' the chromatin~\cite{Brackley2020JPCM}. It would be interesting in the future to study the phase diagram of ``good bridgers'' (e.g. simple spheres). We have also shown previously that cluster formation via the BIA can be dramatically altered by non-equilibrium chemical reactions which stochastically switch the proteins back and forward between a binding and non-binding state [modelling post-translational modifications~\cite{Brackley2017}]. Without switching, BIA clusters coarsen and merge until there is a single cluster. Switching arrests coarsening, leading to multiple small clusters where the constituent proteins exchange with the soluble pool at a rate determined by the switching rate. Active reactions have similarly been shown to arrest coarsening in LLPS~\cite{Zwicker2015}. Such non-equilibrium processes could provide the cell with a means to control droplet formation and size which does not require precise parameter tuning, allowing  fast exchange with a soluble pool but slow internal mixing. It would therefore be of interest in the future to study the interplay between the BIPS and LLPS in a non-equilibrium context.

\small

\subsection*{Methods}

In our simulations we model chromatin as a chain of beads, each of diameter  $\sigma\!=\!10$~nm, representing roughly 1kbp of DNA or 4-5 nucleosomes. The beads interact sterically with each other via a Weeks-Chandler-Anderson (WCA) potential, and they are connected via finitely-extensible non-linear elastic (FENE) springs; a Kratky-Porod potential provides bending rigidity. HP1 dimers are represented as a rigid body made up from 7 spheres of diameter $0.5\sigma$. Steric interactions between HP1 spheres and between HP1 and chromatin beads are provided by a WCA potential. Attractive interactions between particular protein domains, and between protein and chromatin beads are provided by a potential with a functional form similar to the Morse potential. All components are confined in a cube of size 35$\sigma$. Full details of the HP1 geometry and all interaction potentials are given in \SI{}. We perform Langevin dynamics simulations using the LAMMPS molecular dynamics software~\cite{Plimpton1995}. The dynamics are integrated using a velocity-Verlet algorithm with a time step of $0.001\tau$, where $\tau$ is a simulation time unit. All results are obtained by averaging of at least 4 simulations of at least $5\!\times\!10^3\tau$. Further details are given in \SI{}.

We acknowledge support from European Research Council (CoG 648050, THREEDCELLPHYSICS).


\balance

\end{document}


\twocolumn[
  \begin{@twocolumnfalse}
\maketitle

\vspace{0.5cm}

\begin{center}
\Huge\bf Supplementary Material
\end{center}

\thispagestyle{fancy}

\vspace{1cm}

\setcounter{tocdepth}{2}
\tableofcontents

  \end{@twocolumnfalse}
]

\twocolumn[{

\vspace{2cm}

\begin{center}
  \includegraphics{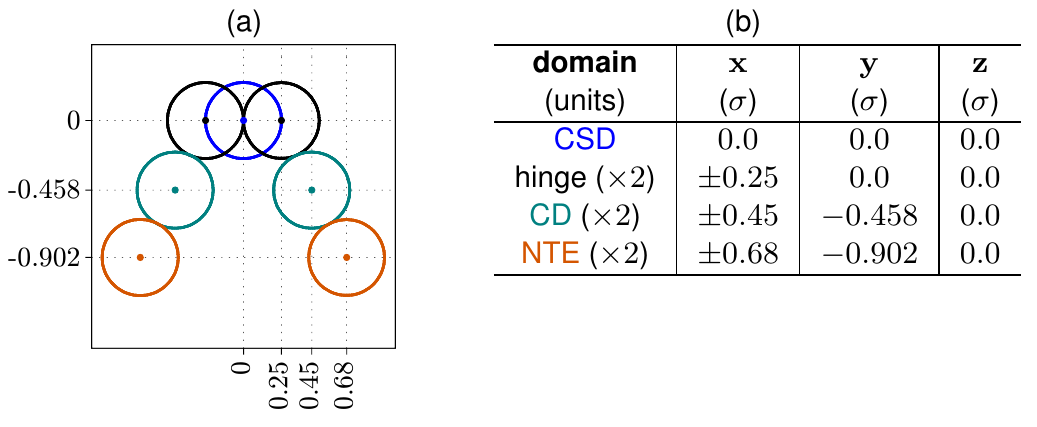}
\end{center}
\captionof{figure}{\textbf{Model HP1 rigid body structure.} (a) Diagram showing the relative positions of the 7 component beads of the model HP1 dimer.
A bead positioned at the origin represents the chromoshadow domain (CSD, blue), which is the dimerisation domain. Other beads represent the two copies (one per monomer) of the hinge domain (black), the chromodomain (green) and the N-terminal end (orange). The same colour scheme is used in \mainFig{1} in the main text; units are chromatin bead diameters $\sigma$, and all HP1 beads have diameter $0.5\sigma$. (b) Coordinates relative to the CSD.
    \label{fig:hp1structure}}

\vspace{3cm}
    
\begin{center}
  \includegraphics{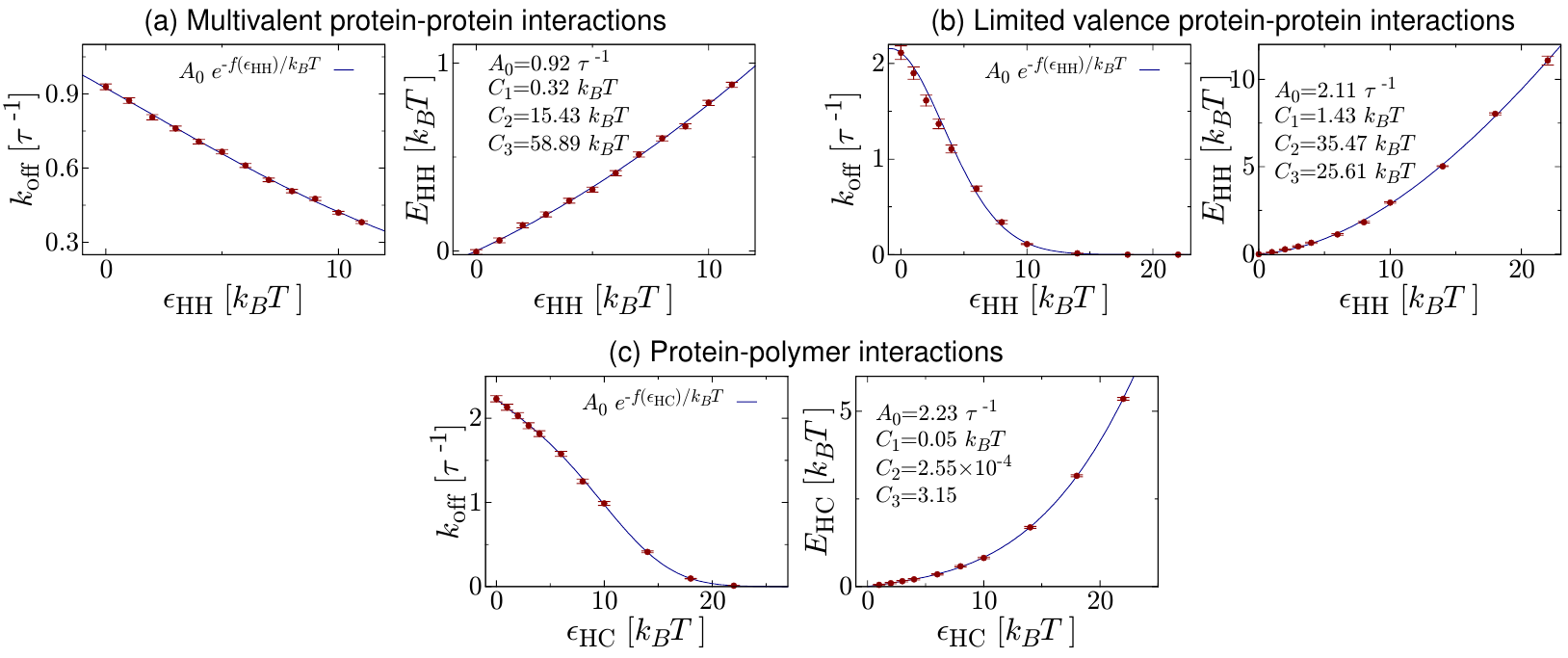}
\end{center}
\captionof{figure}{\textbf{Interaction energy calibration.} Left-hand plots show the mean duration of binding events as a function of the bare interaction energies $\epsilon_{\rm HH}$ or $\epsilon_{\rm HC}$ obtained from calibration simulations as detailed in \SIapp{sec:energycalib}. Points show an average over at least 1000 binding events, and error bars show the standard error in the mean. Lines are obtained from a fit to the data as detailed in \SIapp{sec:energycalib}. Right-hand plots show the mapping between the effective ($E$) and bare ($\epsilon$) interaction energies obtained from this fit, with the values of the fit parameters indicated. Panels (a) and (b) show results for protein-protein interactions from the multivalent and limited valence HP1 models respectively. Panel (c) shows results for HP1-chromatin interactions. \label{fig:energycalib}}
    
}]

\twocolumn[{

\vspace{5cm}
    
    \begin{center}
      \includegraphics{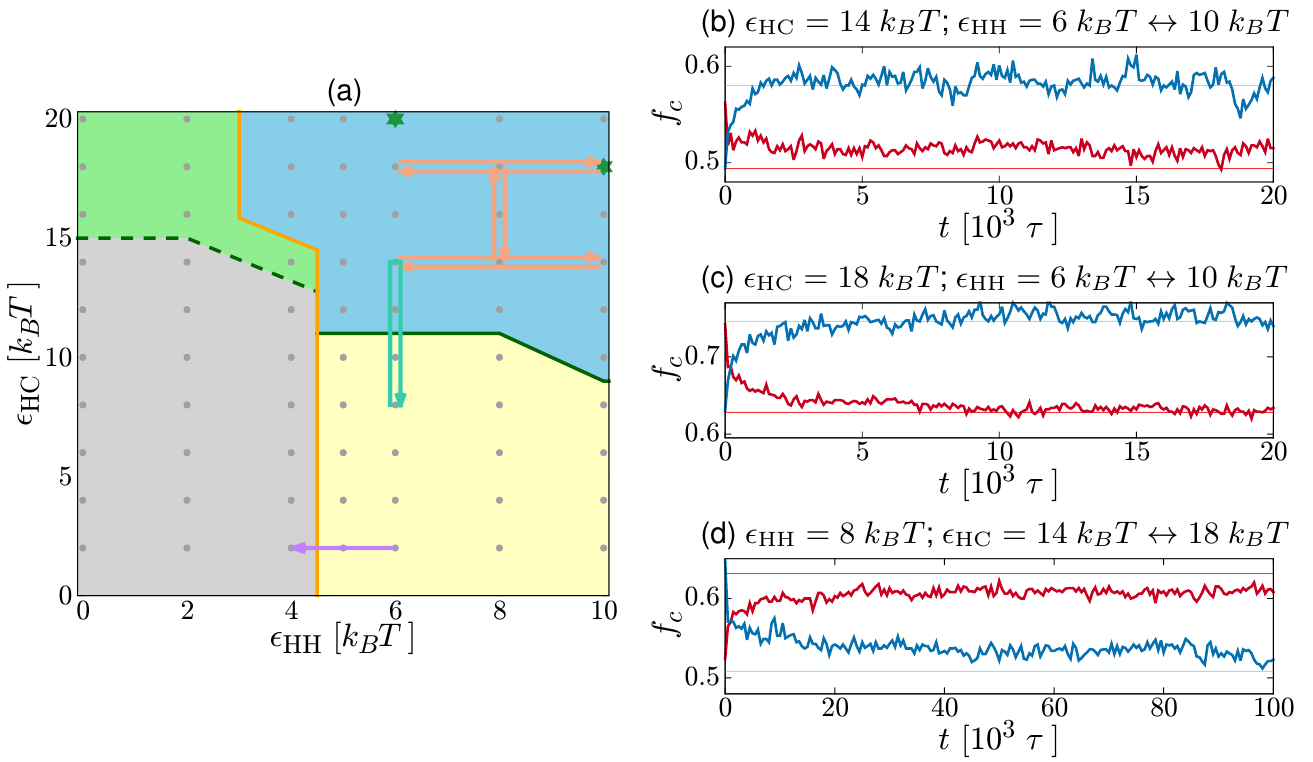}
    \end{center}
    \captionof{figure}{\textbf{Simulated configurations are representative of equilibrium.} (a) Phase diagram for multivalent model. Grey points show parameter values for simulations used in \mainFigs{2-4} in the main text. Peach arrows show quench simulations where after obtaining an equilibrium configuration for one set of parameters, the energy values were instantaneously changed. The dark-green stars indicate points where replica exchange simulations were performed (see \SIapp{sec:eqm}). For comparison, the light-green arrow shows the parameter values used in the hysteresis simulations detailed in \SIapp{sec:hyst} (also \mainFig{3(c)} in the main text). The purple arrow indicates parameters for a simulation where $\epsilon_{\rm HC}$ was instantaneously reduced from $6k_BT$ to $4k_BT$, and we observed that the droplet dissolves (see \SIapp{sec:eqm}). On the right-hand plots we show the number of chromatin beads bound to HP1s, $f_c$, as a function of time after a quench, where the energy values are changed instantaneously. In (b), $\epsilon_{\rm HC}=14k_BT$ and $\epsilon_{\rm HH}$ is changed from $6k_BT$ to $10k_BT$ at $t=0$ (dark-red line), or from $10k_BT$ to $6k_BT$ at $t=0$ (dark-blue line). Pale coloured lines indicate the value of $f_c$ at equilibrium. Panels (c) and (d) show similar plots as indicated.
      \label{fig:quenches}}

    }]

\twocolumn[{

\vspace{0.2cm}

    \begin{center}
      \includegraphics{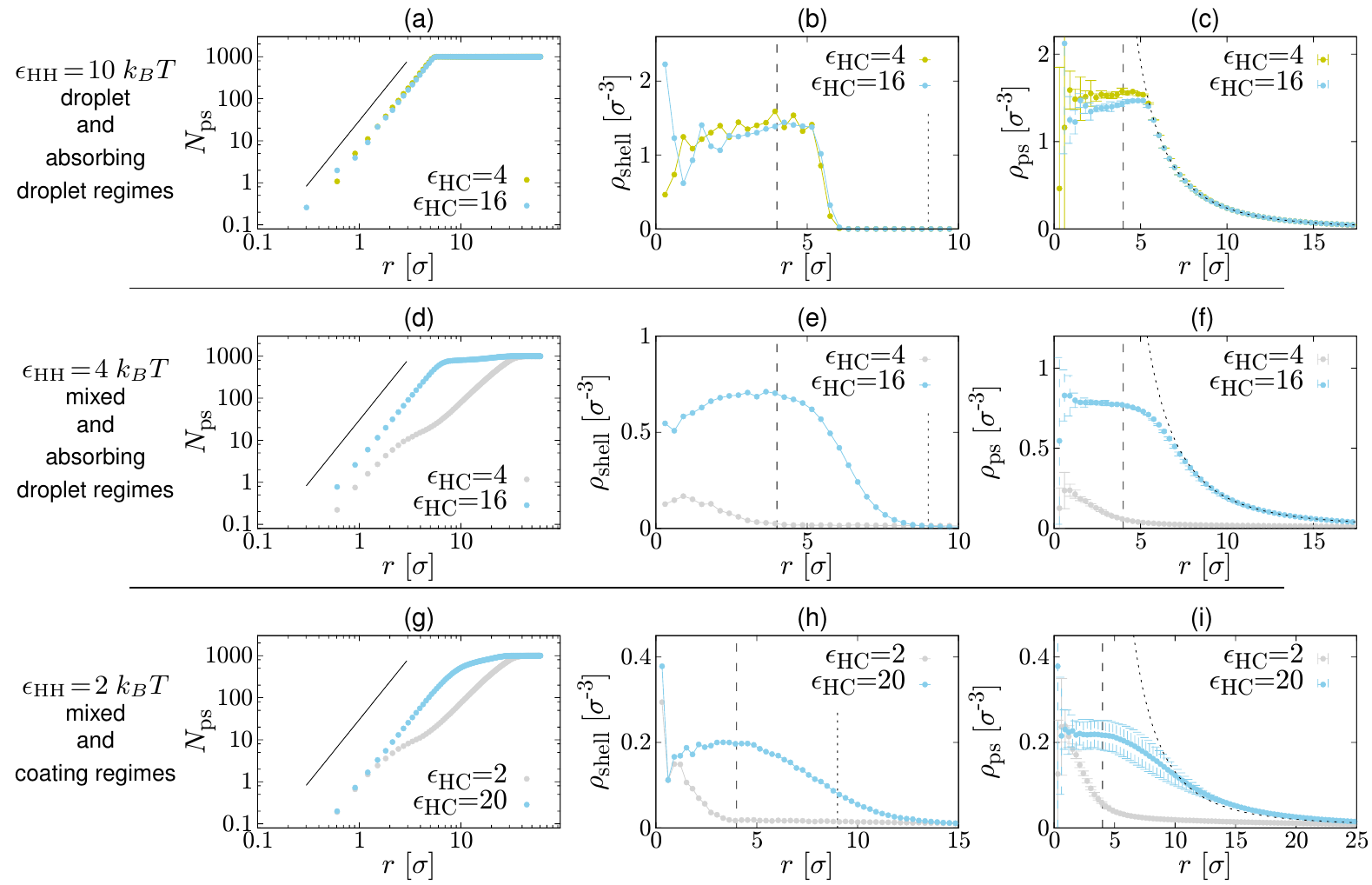}
    \end{center}
    \captionof{figure}{\textbf{Protein densities within and outside protein droplets for the multivalent HP1 model.} Each row shows plots obtained for a different value of $\epsilon_{\rm HH}$ with the relevant regimes indicated on the left. (a) Plot showing how the number of proteins $N_{\rm ps}$ within a probing sphere varies with the sphere radius $r$ (log-log scale). The probing sphere is centred on the centre of mass of the largest protein cluster in the system. Results for two different values of $\epsilon_{\rm HC}$ are shown as indicated (units are $k_BT$). Black lines shows $N_{\rm ps}\sim r^3$, the expected growth for a uniform protein droplet.  (b) The density of proteins within a spherical shell of width $dr=0.3\sigma$ is plotted as a function of $r$. The abrupt drop-off can be used to extract an estimate of the droplet radius. The dashed and dotted vertical lines correspond to $r_{\rm in}$ and $r_{\rm out}$, respectively (see \SIapp{sec:measureDensity}). (c) Here, the mean density within the entire probing sphere is plotted as a function of $r$. We estimate the droplet density $\rho_{\rm HD}$ by fitting a horizontal line to the region between $\sigma$ and $r_{\rm in}$. The black dotted curve represents the expected decay of the density $\rho_{\rm ps}\sim r^{-3}$ outside the droplet. (d-f) Similar plots but for $\epsilon_{\rm HH}=4k_BT$. Our probing sphere procedure can be applied even in the mixed phase, but it does not make sense to extract a density in this case. For the $\epsilon_{\rm HC}=16k_BT$ case in the absorbing
droplet regime there is a slower drop-off in $\rho_{\rm shell}$ than in panel (c), indicating a broader boundary region (where proteins transiently coat polymer segments which extend out of the droplet). Panels (g-i) show similar plots for $\epsilon_{\rm HH}=2k_BT$.
      \label{fig:density}}

 \vspace{1.5cm}
    
	 \begin{center}
      \includegraphics{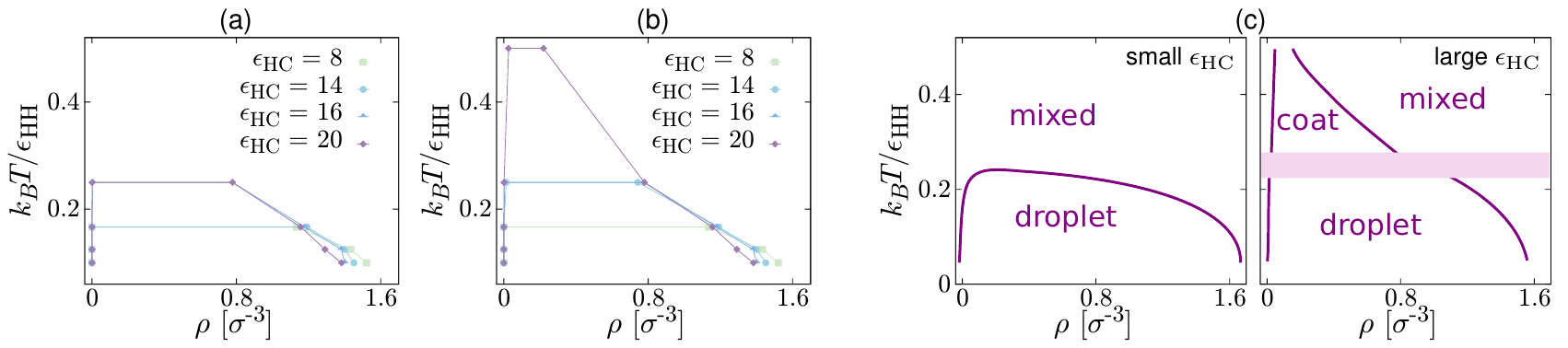}
    \end{center}
    \captionof{figure}{\textbf{Phase densities $\boldsymbol{\rho}_{\rm \mathbf{LD}}$ and $\boldsymbol{\rho}_{\rm \mathbf{HD}}$ for the multivalent HP1 model.} (a) Plot showing values for the protein densities in the low and high density phases, on the $\rho$-$k_BT/\epsilon_{\rm HH}$ plane. Colour indicates the value of $\epsilon_{\rm HC}$ as indicated; points are only shown for parameter values where $\phi_{\rm sep}>0.6$. Connecting lines are shown as a guide to the eye, and give approximate boundaries between mixed and droplet regimes. (b) Similar plot, but now points are shown for all parameter values where $\phi_{\rm sep}>0.2$. This means that there are additional points compared to panel a for large $\epsilon_{\rm HC}$ values; the lines locate the approximate boundaries between mixed and droplet/coating regimes. (c) Sketch phase diagrams for low (left) and high (right) values of $\epsilon_{\rm HC}$ as interpreted from the data in panels (a) and (b).
        \label{fig:denPhaseDiag}}

}]

\twocolumn[{

\vspace{0.5cm}

     \begin{center}
       \includegraphics{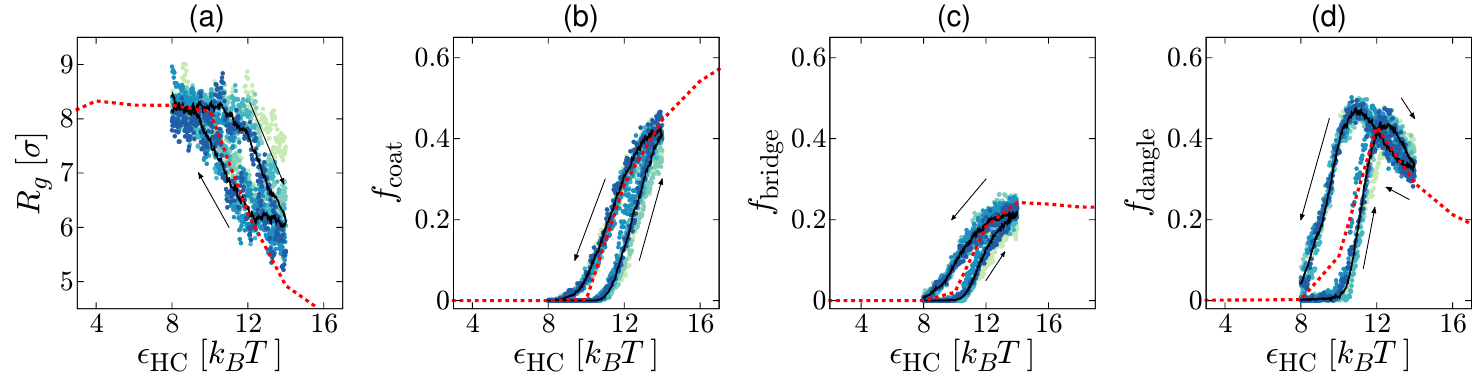}
     \end{center}
     \captionof{figure}{\textbf{Hysteresis in the `droplet'--`absorbing droplet' transition.} (a) Plot showing the polymer radius of gyration obtained from simulations where $\epsilon_{\rm HC}$ is slowly increased from $8k_BT$ to $14k_BT$ before being decreased again as detailed in \SIapp{sec:hyst}. Points show values obtained from 12 individual simulations, with each simulation shown in a different colour. The black line shows an average over these repeat simulations, and arrows indicate the direction of time. The red dotted line shows the equilibrium curve [as in \mainFig{4(d)} in the main text]. (b-d) Similar plot but showing the fraction (of the $N=1000$ proteins) which are bound to the polymer in each of the different binding modes. \label{fig:HyRg}}

\vspace{2cm}
     
    \begin{center}
      \includegraphics{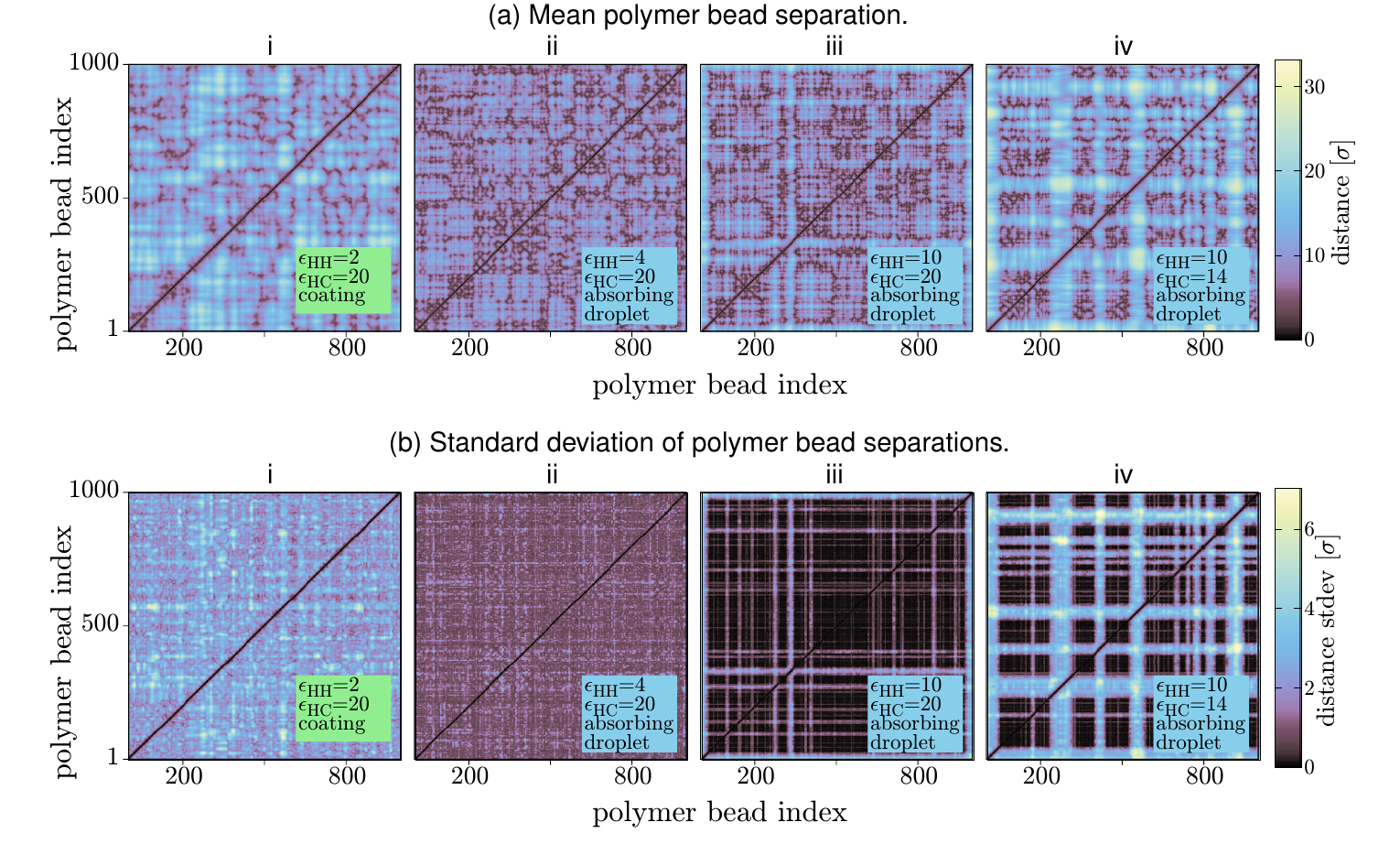}
      \end{center}
    \captionof{figure}{\textbf{Polymer `distance maps' for the multivalent protein model.} (a) Colour maps showing distances between chromatin beads obtained from a time average of single equilibrium simulation (of duration $5\times10^3\tau$). The different sub-panels show maps from the different regimes with $\epsilon_{\rm HH}$ and $\epsilon_{\rm HC}$ as indicated (units of $k_BT$). Distant pairs of chromatin beads (swollen polymer) are shown by brighter colours, while pairs of beads which are close together in 3D space by darker colours (compacted polymer). For example, in map iv, dark and light stripes indicate that swollen and crumpled polymer regions coexist. (b) Similar maps obtained from the same simulations, but showing the standard deviation of the bead separations rather than the mean. Each map is obtained from a single simulation. This gives an indication of how dynamic different polymer regions are, with brighter colours indicating that the separation of the pair of beads varies during the simulation.
\label{fig:CM}}

    }]

\twocolumn[{
    
\vspace{0.3cm}
    
\centering
\includegraphics{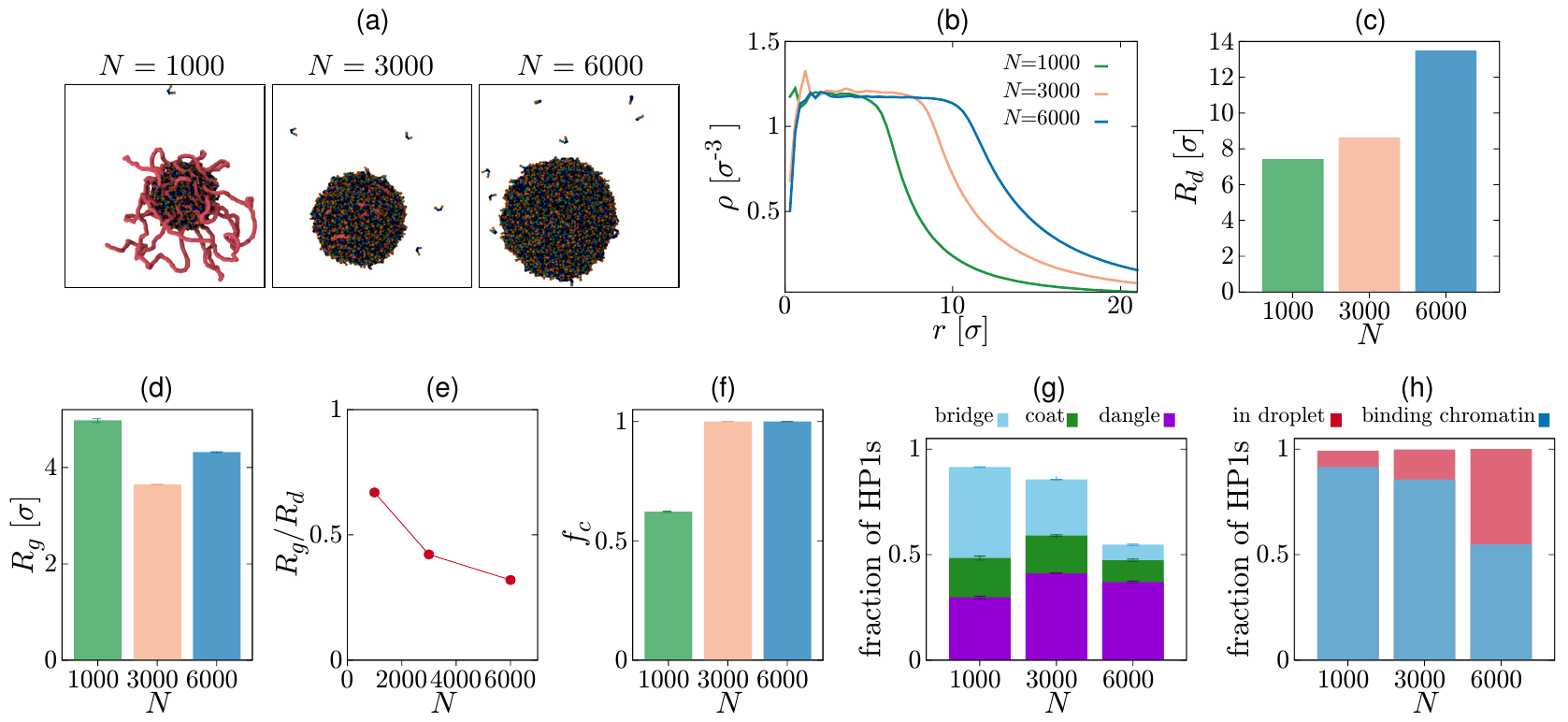}
          \captionof{figure}{\textbf{Varying the number of proteins for parameters where a droplet would also form without chromatin.} Plots showing the effect of varying the number of proteins for parameter values $\epsilon_{\rm HH}\!=\!6k_BT$, $\epsilon_{\rm HC}\!=\!14k_BT$ within the absorbing droplet phase. (a) Snapshots from simulations with $N=1000$, 3000 and 6000 proteins. (b) Density of proteins within a probing sphere of radius $r$ centred on the the centre of mass of the droplet. (c) Bar plot showing the radius of the protein droplet in simulations with different values of $N$. (d) Bar plot showing the radius of gyration of the polymer. (e) Plot showing how the ratio $R_g/R_d$ varies with $N$. (f) Bar plot showing the fraction of polymer beads bound by proteins. (g) Bar plot showing the fraction of the total number of proteins which are bound to the polymer in each of the three modes. Bars are stacked on top of each other so, for example, the distance between the bottom and top of the green region gives the fraction of proteins bound in the coating mode. The total height shows the total fraction of proteins bound to the polymer. (h) Stacked bar plot showing the fraction of the total number of proteins which are in the droplet but not binding to chromatin beads (red) and in the droplet \textit{and} binding to chromatin beads (blue).
	    \label{fig:varyN}}

\vspace{1.5cm}
          
\includegraphics{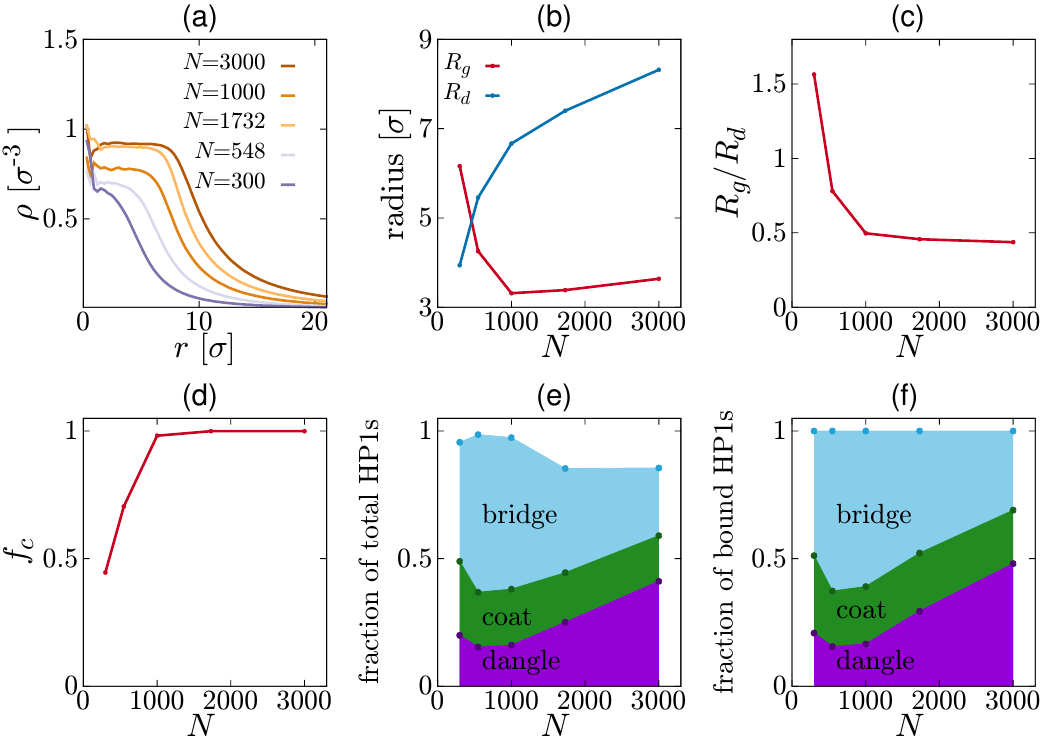}
          \captionof{figure}{\textbf{Varying the number of proteins for parameters where a droplet \textit{only} forms in the presence of chromatin.} Plots showing the effect of varying the number of proteins $N$ for parameter values $\epsilon_{\rm HH}=4k_BT$, $\epsilon_{\rm HC}=20k_BT$ within the absorbing droplet phase, in the regime where phase separation would not occur in the absence of chromatin.  (a) The density of proteins is measured within a probing sphere of radius $r$ centred on the centre of mass of the protein droplet. (b) Plot showing the radius of the droplet and the radius of gyration of the polymer as a function of $N$ [the same plot is shown with a log-scale in \mainFig{4(g)} in the main text]. (c) Plot showing how the ratio $R_g/R_d$ varies with $N$. (d) Plot showing the fraction of polymer beads bound by proteins as a function of $N$. (e) The fraction of proteins which are bound to the polymer in the bridging, coating and dangling modes are shown as a function of $N$. Curves are stacked on top of each other, so for example, the distance between the bottom and top of the green region gives the fraction of proteins bound in the coating mode. (f) A similar plot shows bridging, coating and dangling proteins as a fraction of the number of proteins which are bound to the polymer.
	    \label{fig:varyN2}}
          
    }]

\twocolumn[{

\vspace{1.75cm}

    \begin{center}
      \includegraphics{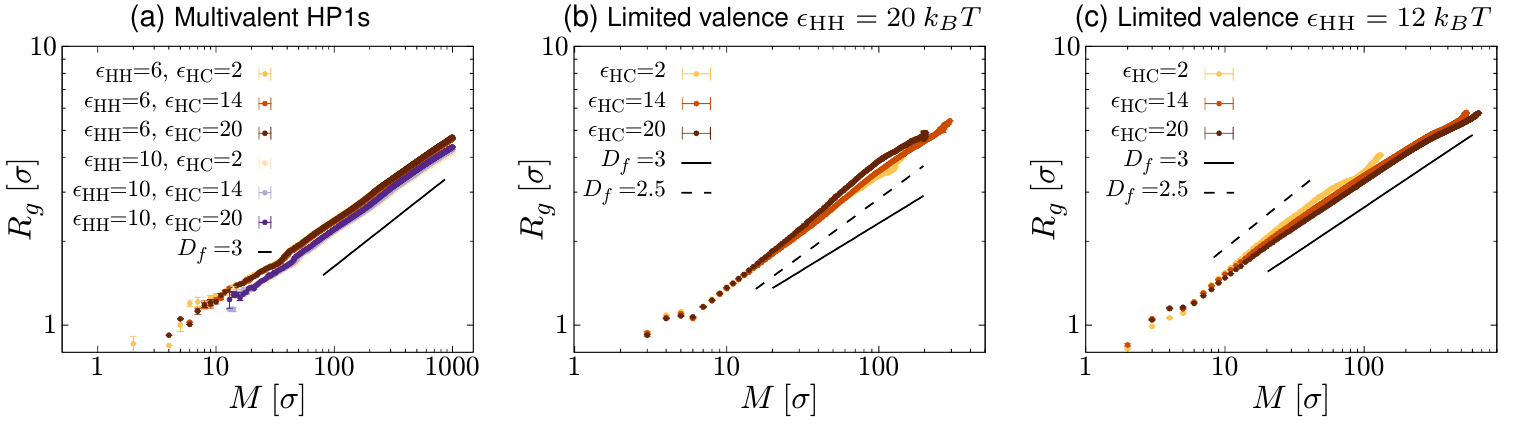}
    \end{center}
\captionof{figure}{\textbf{Fractal dimension of HP1 sub-clusters} Logarithmic scale plots showing radius of gyration $R_g$ \textit{vs.} the number of proteins $M$ in sub-clusters for multivalent and limited valence simulations in droplet/cluster regimes [as detailed in \SIapp{sec:fractal}]. Each point shows the mean $R_g$ of all unique sub-clusters with a given $M$; averages are also over time for a single simulation. Error bars show the standard error in the mean. We expect a power-law relationship with $R_g\sim M^{1/D_f}$, where $D_f$ is the fractal dimension. (a) Points show data obtained from simulations of the multivalent model in the droplet or absorbing droplet regimes, with $\epsilon_{\rm HH}$ and $\epsilon_{\rm HC}$ as indicated (units are $k_BT$). Lines show the slope for the indicated values of $D_f$ (b) Points show data obtained from simulations of the limited valence model with large $\epsilon_{\rm HH}=20k_BT$ and $\epsilon_{\rm HC}$ as indicated (clustering regimes). (c) Points show data from simulations of the limited valence model with smaller $\epsilon_{\rm HH}=12k_BT$.  \label{fig:fractal}}

\vspace{3cm}

\begin{center}
  \includegraphics{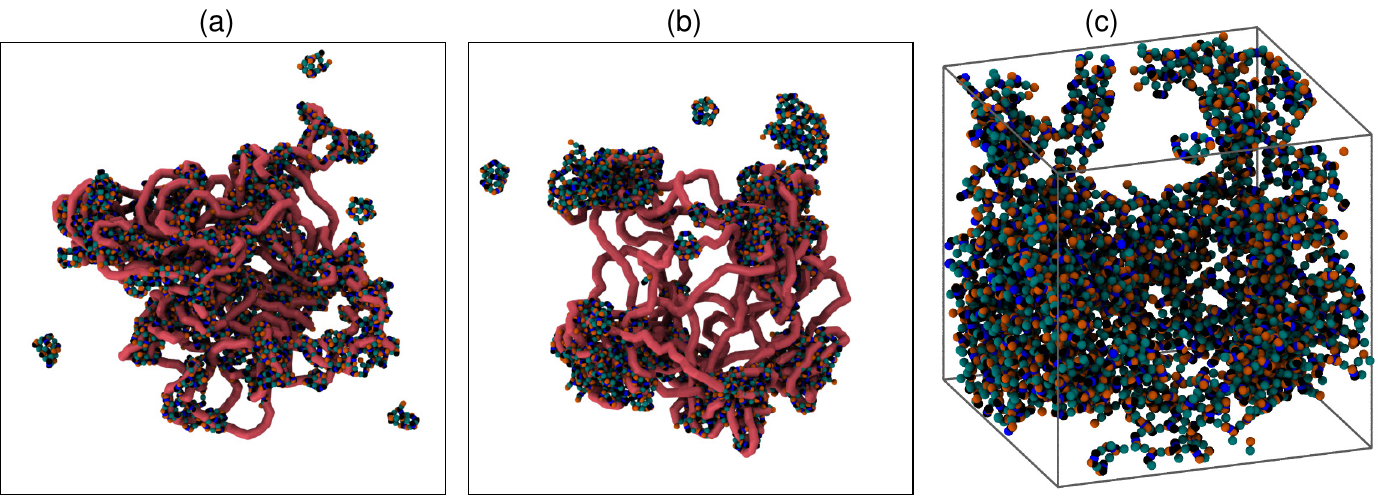}
\end{center}
\captionof{figure}{\textbf{Alternative quench schemes lead to different long-lived metastable configurations for limited valence model proteins.} Snapshots are shown from simulations of the limited valence model using parameters $\epsilon_{\rm HH}=16k_BT$ and $\epsilon_{\rm HC}=16k_BT$. (a) Configuration obtained from the end of a $2\times10^4\tau$ long simulation where for the first $10^4\tau$ the HP1-HP1 attraction was switched off; the HP1-chromatin attraction was kept switched on for the full duration. (b) Configuration obtained from the end of a $2\times10^4\tau$ simulation where for the first $10^4\tau$ the HP1-chromatin  attraction was switched off and instead the HP1-HP1 attraction was on for the full duration. (c) Configuration from a simulation of duration $4\times10^3\tau$ with a higher density of HP1s in the absence of polymer ($N=1000$ proteins in a smaller system of side $l_x=21\sigma$) using periodic boundary conditions (instead of `walls' as in all other simulations). \label{fig:snapalt}}

    }]

\twocolumn[{

\begin{center}
  \includegraphics{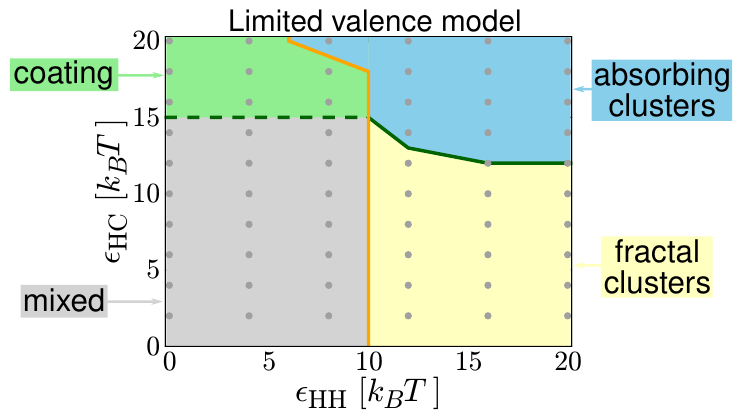}
\end{center}
\captionof{figure}{\textbf{Phase diagram for limited valence model HP1s.} Phase diagram showing the different behaviours of the limited valence HP1s with different parameter values. Colours correspond to those used in the equivalent regimes for the multivalent case shown in \mainFig{2} in the main text. Grey points indicate parameters used in different simulations. The position of the orange line is determined using $\phi_{\rm sep}=0.5$ as a threshold, while the position of the solid green line is determined by the total fraction of proteins bound to the polymer $f_{\rm tot}$ (we define the absorbing clusters regime as where $\phi_{\rm sep}\geq0.5$ and $f_{\rm tot}\geq0.5$). As with the multivalent case, $\phi_{\rm sep}$ is approximately independent of $\epsilon_{\rm HC}$ in the mixed phase, and we set the position of the dashed green line as the point where $\phi_{\rm sep}$ first starts to increase with $\epsilon_{\rm HC}$. \label{fig:ph_diag_LV}}

\vspace{1.7cm}

\begin{center}  
  \includegraphics{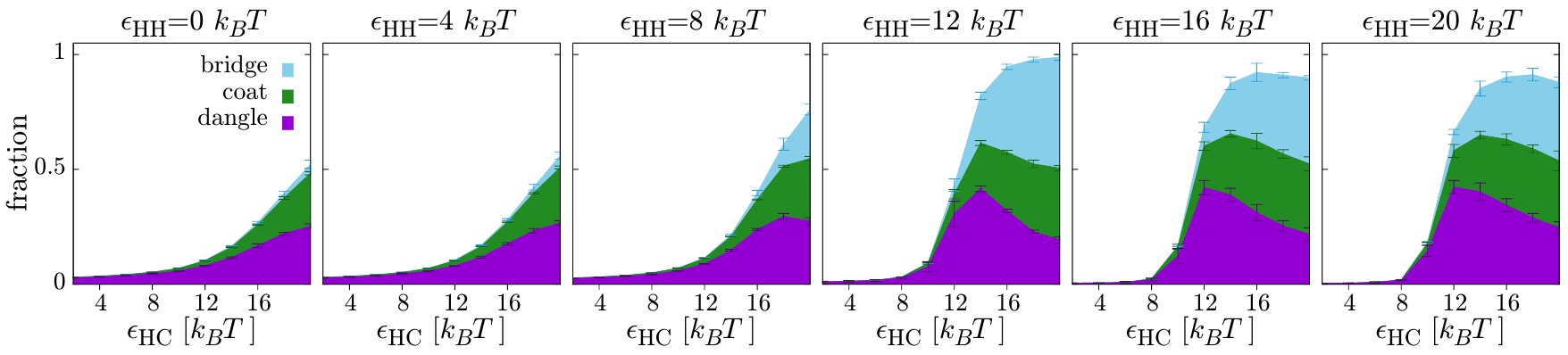}
\end{center}
\captionof{figure}{\textbf{Protein-chromatin binding modes for the limited valence protein model.} Plots showing the fraction of the $N=1000$ proteins bound to the chromatin in each mode for different interaction energies. The height of each coloured region indicates the proportion of proteins, with the regions stacked on top of each other. In this way the height of the total coloured region indicates the fraction of proteins bound in any mode $f_{\rm tot}$. Values are obtained from averaging over 4 simulations of duration $10^4 \tau$, and error bars show the standard error of the mean.\label{fig:LVmodes}}

\vspace{1.7cm}

\begin{center}
  \includegraphics{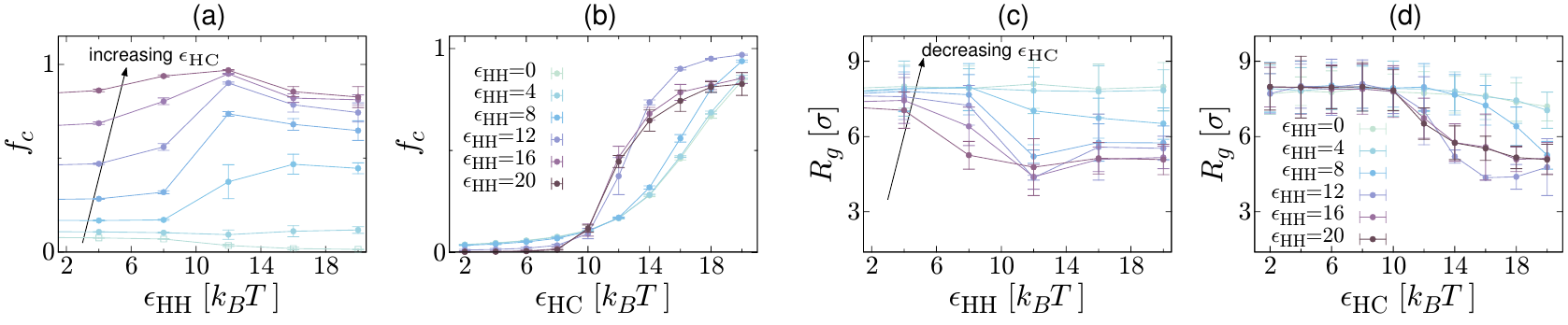}
\end{center}
\captionof{figure}{\textbf{HP1-chromatin interactions and compaction for the limited valence protein model.} (a-b) Plots showing how the fraction of chromatin beads which are bound by proteins $f_c$, depends on the interaction energies. In (a) from bottom to top curves are for $\epsilon_{\rm HC}$ between 8 and 20$k_BT$ increasing in steps of $2k_BT$. Points are obtained from an average of $4$ independent simulations of equilibrium configurations; error bars show the standard error of the mean, and connecting lines are drawn as a guide to the eye. (c-d) Plots showing how the radius of gyration of the polymer representing the chromatin segment depends on the interaction energies. In (c), from top to bottom, curves are again for $\epsilon_{\rm HC}$ between 8 and 20~$k_BT$ increasing in steps of 2$k_BT$. \label{fig:Rg}}

    }]

\twocolumn[{

\vspace{5.5cm}

\begin{center}
  \includegraphics{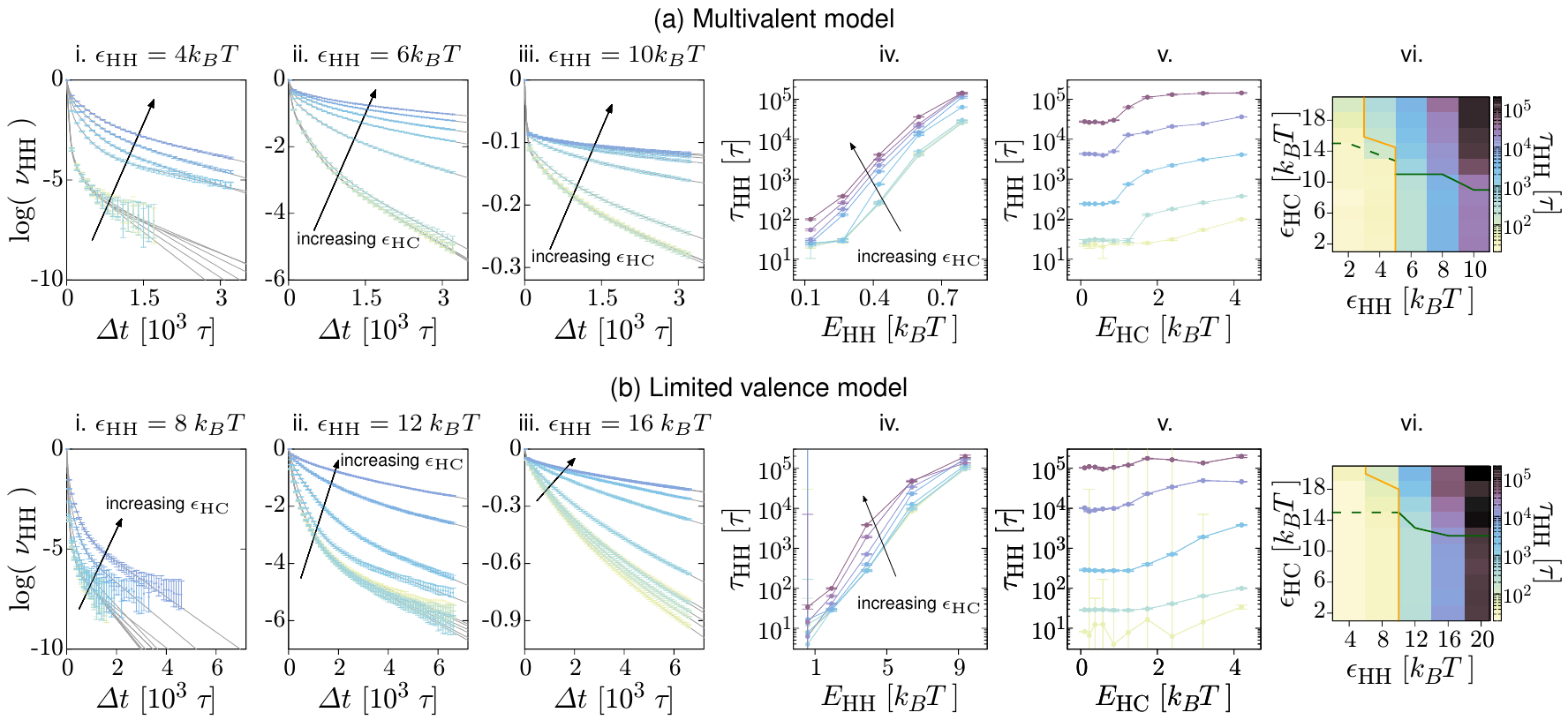}
\end{center}
\captionof{figure}{\textbf{Protein dynamics measurements}. Results for (a) multivalent model proteins, and (b) limited valence model proteins. i-iii. Plots showing $\nu_{\rm HH}$ as a function of the time interval $\Delta t$ on a log-linear scale. Coloured points are obtained from 4 independent simulations with error bars showing the standard error. Grey lines show fits to Eq.~(\ref{eq:3times}). Where $\nu_{\rm HH}$ gets very small [e.g., in (a)i] the errors can become quite large; we do not show points where the fractional error is larger than 1. iv-v. Plots showing the values of $\tau_{\rm HH}$ as extracted from the $\nu_{\rm HH}(\Delta t)$ fits (see \SIapp{sec:dynamics}) at different energy parameter values. Here we plot against the effective energies $E_{\rm HH}$ and $E_{\rm HC}$, as defined in \SIapp{sec:energycalib}. We note that the relationships look very similar to those found for $\epsilon_{\rm HH}$ and $\epsilon_{\rm HC}$ as plotted in \mainFig{6} in the main text. Points with error bars are obtained from averages over 4 independent simulations, while lines are added as a guide to the eye. vi. Colour maps showing how $\tau_{\rm HH}$ varies with $\epsilon_{\rm HH}$ and $\epsilon_{\rm HC}$. Lines separating the various regimes are overlaid as in \mainFig{2(b)} in the main text and \SIFig{fig:ph_diag_LV}. 
\label{fig:dynamics}}

}]

\clearpage

\clearpage

\section{Chromatin model}

In our coarse grained scheme the chromatin fibre is modelled as a chain of $L$ spherical beads of diameter $\sigma$, and the position of the $i$th bead is denoted by $\mathbf{r}_i$. Consecutive beads along the chain are connected by finitely extensible non-linear elastic (FENE) bonds described by 
\begin{equation}
\begin{split}
U_{\rm FENE}(r_{i,i+1}) = & \; U_{\rm WCA}(r_{i,i+1}) \, + \\
 &-\frac{K_{\rm FENE} R_0^2}{2} \log \left[ 1 - \left(\frac{r_{i,i+1}}{R_0}\right)^2 \right],
\end{split}
\label{eq:FENE}
\end{equation}
where $r_{i,i+1} = |\mathbf{r}_{i+1}-\mathbf{r}_i|$ is the bead separation, and we set the bond energy $K_{\rm FENE} = 30 k_{\rm B}T$, and maximum extension $R_0 = 1.6 \sigma$. The Boltzmann constant and the temperature are denoted $k_B$ and $T$ respectively, and we use energy units of $k_BT$ throughout. The first term in Eq.~(\ref{eq:FENE}) is the Weeks-Chandler-Andersen (WCA) potential
\begin{align}
\frac{U_{\rm WCA}(r_{ij})}{k_BT}  = \left\{ 
\begin{array}{ll} 
4 \left[ \left( \frac{d_{ij}}{r_{ij}}\right)^{12} - \left( \frac{d_{ij}}{r_{ij}}\right)^{6} \right] + 1 & r_{ij}<2^{1/6}d_{ij}, \\
0 & \mbox{otherwise},
\end{array} \right.
\label{eq:WCA}
\end{align}
where $d_{ij}$ is the mean of the diameters of the beads $i$ and $j$ (i.e., $d_{ij}=\sigma$ for $i,j\in 1 \ldots L$). This gives a purely steric interaction preventing beads from overlapping. Non-adjacent beads also interact via the WCA. 

To model the bending rigidity of the fibre, the Kratky-Porod potential is introduced between triplets of adjacent beads
\begin{equation}
U_{\rm BEND}(\theta_i)=\\K_{\rm BEND} \left[ 1 - \cos(\theta_i) \right],
\label{eq:bend}
\end{equation} 
where $\theta_i$ is the angle formed between beads $i-1$, $i$ and $i+1$, as given by
\begin{equation}
\cos(\theta_i) = \frac{[\mathbf{r}_i-\mathbf{r}_{i-1}]\cdot [\mathbf{r}_{i+1}-\mathbf{r}_{i}]}{|[\mathbf{r}_i-\mathbf{r}_{i-1}]\cdot [\mathbf{r}_{i+1}-\mathbf{r}_{i}]|},
\end{equation}
and $K_{\rm BEND}$ is the bending energy. The persistence length in units of $\sigma$ is given by $l_p=K_{\rm BEND}/k_BT$, and we set $l_p=4\sigma$. A typical mapping of the polymer model to a chromosome fragment is to consider one bead to represent approximately 10~nm (about 1kbp or $4$--$5$ nucleosomes) of chromatin; then $l_p\simeq120$~nm, which is reasonable for chromatin~\cite{Brackley2013}.

\section{HP1 dimer models}

As detailed in the main text, each HP1 dimer is represented by a rigid body consisting of seven spheres arranged as shown in \mainFig{1(a)}: the blue sphere represents the CSD from each HP1 making up the dimer (i.e., two CSDs), then the black, green and orange spheres represent the hinge, CD and NTE domains respectively, two of each per dimer. Each sphere has diameter $0.5 \sigma$, with position coordinates relative to the CSD sphere as shown in \SIFig{fig:hp1structure}.

Interactions between HP1 component beads and chromatin beads are modelled as follows. CDs interact attractively with chromatin beads through the potential
\begin{equation}
  U_{\rm CD-C}(r)= \left\{
    \begin{array}{ll}
      \epsilon_{\rm HC} \left[ (\rm{e}^{-2\alpha r} - 2\rm{e}^{-\alpha r}) \right. & ~ \\
     \;\quad\quad  \left. - (\rm{e}^{-2\alpha r_{\rm HC}} -2\rm{e}^{-\alpha r_{\rm HC}}) \right] & r \leq r_{\rm HC}, \\
      0 & \mbox{otherwise},
\end{array}
      \right.
\label{eq:CD-C}
\end{equation} 
where $r$ is the separation between the centres of the CD and the chromatin bead, $\epsilon_{\rm HC}$ is the energy which determines the strength of the interaction, $\alpha$ is a ``shape'' parameter, and $r_{\rm HC}$ is the cut-off distance which sets the range of the interaction. We set $\alpha = 5$ and $r_{\rm HC} = 0.9\sigma$. 
The CSD, hinge, and NTE interact sterically with the chromatin beads, through the WCA potential given in~\eqref{eq:WCA}; we set $d_{ij}=0.75\sigma$ for the CSD and $d_{ij}=0.5\sigma$ for the hinge and NTE. This allows partial overlap of these beads with the chromatin beads, which is essential to permit interaction between the CDs and chromatin beads (it also accounts for the hinge and NTE being flexible/disordered domains).

Interactions between HP1s are modelled similarly. The hinge and NTE domains in different HP1 dimers interact via the potential 
\begin{equation}
U_{\rm h-NTE}(r)= \left\{
\begin{array}{ll}
  \epsilon_{\rm HH} \left[ (\rm{e}^{-2\alpha r} -2\rm{e}^{-\alpha r}) \right. & ~\\
    \;\quad\quad \left. - (\rm{e}^{-2\alpha r_{\rm HH}} -2\rm{e}^{-\alpha r_{\rm HH}}) \right] & r \leq r_{\rm HH}, \\
  0 & \mbox{otherwise}.
  \end{array} \right.
\label{eq:h-NTE}
\end{equation}
The two different models of the HP1-HP1 interaction are specified by different sets of parameters: for the \textit{multivalent} model we set $\alpha=0.5$ and $r_{\rm HH}=1.3\sigma$; for the \textit{limited valence} model we set $\alpha=5$ and $r_{\rm HH}=0.6\sigma$. All other HP1 component beads interact sterically via the WCA, except for the CSD and the NTE which can overlap (there is no interaction) in order to permit the correct binding of NTE domains with hinges.

We note that the minima of the functions $U_{\rm CD-C}(r)$ and $U_{\rm h-NTE}(r)$ is at $r=0$, i.e., the beads can overlap. In practice, it is not always possible to achieve the separation $r=0$ due to steric interactions between the other component beads. For the HP1-chromatin interaction, allowing the CD and chromatin beads to overlap, along with the short range of the interaction, ensures that a given CD can interact with at most one chromatin bead at a time (e.g., a single CD cannot form a bridge between two chromatin beads). This is a reasonable choice, since a chromatin bead represents several nucleosomes, and the CD is thought to interact with nucleosome surface charges. We also note that the functional forms of  $U_{\rm CD-C}(r)$ and $U_{\rm h-NTE}(r)$ are the same as the commonly used Morse potential.

In the \textit{limited valence} case, the short range of $U_{\rm h-NTE}(r)$ with no repulsive core means that one NTE can interact with exactly one CSE at a time and \textit{vice versa}. For the \textit{multivalent} case the longer range and small value of $\alpha$ (which leads to a broader shape) is such that multiple NTEs can simultaneously interact with one CSE and \textit{vice versa}.

\section{Langevin dynamics}

We use the LAMMPS software~\cite{Plimpton1995} to perform Langevin dynamics simulations. Briefly, the position of each polymer bead and the centre of mass of each HP1 dimer is governed by the equation
\begin{equation}
m_i \frac{ d^2 \mathbf{r}_i }{dt^2} = -\nabla U_i - \xi_i \frac{d\mathbf{r}_i }{dt} + \sqrt{2k_BT\xi_i}\boldsymbol{\eta}_i(t),
\label{langevin}
\end{equation}
where $\mathbf{r}_i$ is the position of the centre of mass of the $i$th bead or HP1, $m_i$ is its mass, $U_i$ is the sum of all the interaction potentials for object $i$. The friction $\xi_i$ sets the diffusion constant $D_i=k_BT/\xi_i$ for each object; we set $m_i = 1$ and $\xi_i=0.5$ for polymer beads; for simplicity, we also set the mass of HP1 component beads to $1$, and the friction for the HP1 rigid bodies to $0.5$. The vector $\boldsymbol{\eta}_i(t)$ is a noise term with components which satisfy
\begin{equation}
  \langle \eta_{i\alpha}(t) \rangle = 0 ~~\mbox{and}~~  \langle \eta_{i\alpha}(t)\eta_{j\beta}(t') \rangle = \delta_{ij} \delta_{\alpha\beta} \delta(t-t'),
\end{equation}
where $\eta_{i\alpha}$ is component $\alpha$ of the noise vector for object $i$, and $\delta_{ij}$ and $\delta(t)$ are the Kronecker and Dirac delta functions respectively. The orientation of the HP1 rigid bodies is governed by a similar Langevin equation for rotation. These equations are solved using a velocity-Verlet algorithm with time step dt=0.001$\tau$, where $\tau$ is the simulation time unit defined by $\tau = \sqrt{m \sigma^2/k_BT}$.

One can make a rough mapping between simulation and real time units by considering the typical time for a polymer bead to diffuse across its own diameter $\tau_{\rm Br}=\sigma^2/D$. Then using the Stokes-Einstein relation for a sphere $D=k_BT/(3\pi\eta\sigma)$ and taking the diameter to be $\sigma=10$~nm and the viscosity of nucleoplasm as 10~cP we obtain $\tau\approx0.456$~ms.

\section{Interaction energy scales}\label{sec:energycalib}

In the previous section we introduced $\epsilon_{\rm HC}$ and $\epsilon_{\rm HH}$ as the interaction energies for protein-protein and protein-chromatin interactions respectively, and we use these throughout the main text. We note that due to the form of Eq.~(\ref{eq:h-NTE}) the energy at the minima of this function is not equal to $\epsilon_{\rm HC}$ (the second term shifts the function upwards). The same is true of Eq.~(\ref{eq:CD-C}) and $\epsilon_{\rm HH}$. We further note that while for both functions the minima is at $r=0$, due to the geometry of the protein (and stetic interactions between the different component beads) the separation may not in practice be able to reach zero. We therefore expect the interactions in our model to have some effective strengths $E_{\rm HC}\neq \epsilon_{\rm HC}$ and $E_{\rm HH}\neq \epsilon_{\rm HH}$. It is not guaranteed that the relationship between the effective and ``bare'' interaction energies are linearly related. We therefore measured this empirically from a set of calibration simulations.

A simple assumption is that (in the absence of cooperative interactions) binding events behave as in the Kramer's escape problem~\cite{mel1991kramers}; in Kramer's approximation, the mean duration of binding is $\langle\tau_{\rm bind}\rangle=\tau_0 \rm{e}^{\epsilon/k_BT}$, where $\epsilon$ is the interaction energy.
$\tau_0$ is the typical time for which the two diffusing objects would remain in contact in the absence of an attraction (``in contact'' meaning having separation less than the interaction range). We performed a set of calibration simulations of two HP1 dimers in which one hinge in one dimer was allowed to interact with one NTE in the other. We measured the duration of HP1-HP1 interaction events; \SIFig{fig:energycalib}(a) left panel shows a plot of the unbinding rate $k_{\rm off}=\langle\tau_{\rm bind}^{-1}\rangle$ as a function of $\epsilon_{\rm HH}$ for the multivalent HP1 model. We then fit to obtain the function $E_{\rm HH}=f(\epsilon_{\rm HH})$, which gives a mapping between the bare and effective interaction energies (\SIFig{fig:energycalib}(a) right panel). We considered a number of functional forms for $f(\epsilon)$, finding a good fit for $f(\epsilon)=C_1\epsilon-C_2(1-e^{-\epsilon/C_3})$; best fit values for the constants $C_1$, $C_2$ and $C_3$ are given in the figure. A similar scheme was used in Ref.~\cite{Brackley2020JPCM}. We repeated the same process for the limited valence HP1 model [\SIFig{fig:energycalib}(b)].

To calibrate the protein-polymer interactions we performed a similar set of simulations, but with a short 100 bead polymer and a single HP1 dimer (in which only one CD was allowed to bind to the polymer). We measured the unbinding rate $k_{\rm off}$ as a function of $E_{\rm HC}$ [\SIFig{fig:energycalib}(c) left], and then fit to obtain a function $\epsilon_{\rm HC}=f(E_{\rm HC})$. Here the function $f(E)=c_1E+c_2E^{C_3}$ gives a good fit, leading to the energy mapping in \SIFig{fig:energycalib}(c) right.

In all of the calibration simulations we used a small periodic system size to minimise the time interval between binding events. We used a very long run time (at least $5\times10^6\tau$), and we measured the bead separations at every time step to ensure we properly captured the entire binding event time.

\section{Obtaining equilibrium configurations}\label{sec:eqm}

In this work, for the case of the multivalent HP1 proteins we have focused on the equilibrium properties of the system. A common consideration in molecular dynamics simulations is ensuring that trajectories are representative of the equilibrium state. In order to obtain equilibrium configurations we used a specific annealing process, and also performed a number of additional test simulations.

Initially, we started with a configuration in which the polymer followed the path of a random walk, and proteins were positioned at random within the confinement volume. An initial short simulation using purely repulsive ``soft'' interactions was used to remove bead overlaps; then a longer run was performed using the force field detailed above, but without attractive interactions ($\epsilon_{\rm HC},\epsilon_{\rm HH}=0$). One scheme is then to instantaneously switch on attractive interactions with the desired energy, i.e., performing an instantaneous quench from high to low temperature. Using this scheme, for large $\epsilon_{\rm HH}$ we found that multiple clusters of proteins formed quickly (indicating spinodal decomposition). These clusters proceeded to coarsen via coalescence and Ostwald ripening (while single proteins can escape from a cluster, we did not observe large clusters breaking into smaller ones). Without any mechanism to arrest cluster coarsening we expect to observe a single cluster at equilibrium; however, these dynamics can be very slow (especially diffusion of large protein clusters), and in general a single cluster could not be obtained within a reasonable simulation time. For intermediate values of $\epsilon_{\rm HH}$, after a quench we often observed a single or small number of clusters form, as would be expected from a nucleation and growth process. Obtaining a single cluster was therefore much quicker in this case.

For the majority of our simulations we therefore employed a different scheme designed to obtain equilibrium (single cluster) configurations more quickly. Specifically, we first switched on the protein-polymer interaction (an instantaneous quench from zero to the desired value); then we slowly increased the protein-protein interaction strength over an extended time. The idea being that the system would first move through the parameter regime where the dynamics follow nucleation and growth (and interaction with the polymer would promote nucleation). We then ran the simulation for long enough to obtain a single cluster. In all cases, once a single cluster configuration was achieved we then used that as an initial configuration for another simulation of length $5\times10^3 \tau$, from which we obtained our results. All of the results presented for the multivalent HP1 model are obtained from an average over at least $4$ independent simulations. We also checked that measured quantities (such as $R_g$ and the number of proteins bound to the polymer in the different modes) were fluctuating about steady values, and were not systematically changing during the simulation.

Although the above scheme generates single-cluster configurations, it is still possible for the system to become stuck in a long-lived metastable configuration not representative of equilibrium, particularly if the interaction energies are large. For example, within the absorbing droplet phase---where the fraction of polymer absorbed, $f_c$, depends on the interaction energies---it is important to verify that this really is reflective of equilibrium. 
We performed some additional quench simulations within the absorbing droplet phase, confirming that after a sudden change of parameters $f_c$ relaxes to what we expect is the equilibrium value. In \SIFig{fig:quenches}(a) we show on a phase diagram the various quench simulations performed; \SIFigs{fig:quenches}(b-d) show $f_c$ as a function of time after a quench at $t=0$.
After the quenches at constant values of $\epsilon_{\rm HC}$ [\SIFigs{fig:quenches}(b,c)], we find that $f_c$ relaxes to the expected value within about $5\times10^4\tau$. This verifies that our configurations are representative of equilibrium (measurements of the polymer $R_g$ show similar behaviour). At constant $\epsilon_{\rm HH} = 8k_BT$, after a sudden change in $\epsilon_{\rm HC}$ the relaxation towards the expected value is much slower [\SIFig{fig:quenches}(d)]; however, the continuous decrease of $f_c$ suggests that the equilibrium value would be reached in a longer simulation.

Another important parameter regime is at intermediate $\epsilon_{\rm HH}\approx4~k_BT$, where the droplet only forms if the protein-chromatin interaction is strong enough. To verify that the droplet is indeed unstable for small $\epsilon_{\rm HC}$, we performed a simulation where the initial condition was an equilibrium configuration obtained at $\epsilon_{\rm HH}=6k_BT$, $\epsilon_{\rm HC}=4k_BT$; we then instantaneously reduced the protein-protein interaction to $\epsilon_{\rm HH}=4~k_BT$ [purple arrow in \SIFig{fig:quenches}(a)]. We observed that the protein cluster breaks apart; this rules out the possibility that droplets are stable for these parameters, but just take a very long time to form.

Finally, for two sets of parameter values within the absorbed droplet phase [green stars in \SIFig{fig:quenches}(a)] we sample equilibrium configurations using the replica exchange method~\cite{sindhikara2008exchange}. During a replica exchange simulation (also known as parallel tempering) a set of $N$ independent simulations are performed in parallel, each at a slightly higher temperature than the last, $T_1<T_2<\ldots<T_N$. At regular time intervals, the configurations in simulations $i$ and $i+1$ are exchanged according to a Monte Carlo update rule. The aim is to more easily sample configurations which are rare at lower temperatures, through the exchange of configurations with replicas at higher temperatures; effectively, this allows the system to escape from local free energy minima. Replica exchange is implemented natively in LAMMPS~\cite{Plimpton1995}. For each parameter pair we performed a set of 36 simulations at temperatures between 1.0 and 2.05 (units of $m\sigma^2/k_B \tau^2$). We used the end point of our standard simulations as an initial condition, and ran the replica exchange for $10^4\tau$ (exchanges were attempted every 0.1$\tau$, and 70\% of attempts were successful). The temperature in each replica simulation was then reduced gradually back to 1.0 over a further 500$\tau$ (without further exchanges), before a standard (constant $T$) simulation of length $5\times10^3\tau$ was performed. For both parameter sets, the values of $f_c$, $R_g$ and $f_{\rm tot}$ were consistent across these simulations and with our original simulations. This suggests that our original shorter simulations are indeed representative of equilibrium, and not a long-lived metastable state.

As noted in the main text, the limited valence HP1s behave more like classic patchy particles which are known to exhibit long-lived dynamically arrested non-equilibrium phases including gels and closed loops (which can form at equilibrium at zero temperature)~\cite{russo2009reversible,sciortino2011reversible,bianchi2011patchy,lindquist2016formation}. This means that it is more difficult to obtain true equilibrium configuration that in the multivalent case. For the limited valence model, we therefore did not seek to explicitly obtain equilibrium configurations. Instead we ran each simulation for $2\times10^4\tau$ after starting from a configuration obtained for $\epsilon_{\rm HH},\epsilon_{\rm HC}=0$, taking measurements from the final $10^4\tau$. We then checked that quantities such as $f_c$, $f_{\rm tot}$ and $R_g$ were not systematically changing during this time. Thus, our simulated structures represent a metastable or dynamically arrested state obtained via a rapid quench from low to high interaction energies.

\section{The $\boldsymbol{\rho}-\boldsymbol{\epsilon}_{\rm \mathbf{HH}}$ phase diagram}\label{sec:measureDensity}

We return now to the multivalent HP1 model. In order to calculate the protein densities, e.g., within or outside of a protein droplet, we consider a `probe sphere' of radius $r$ centred on the centre of mass of the largest droplet (protein cluster). We then progressively increase $r$, and calculate the density of proteins within the probe sphere, and within a spherical shell of width $dr$. We then average over time and repeat simulations (finding a new droplet centre of mass each time). 

In \SIFig{fig:density} we plot several quantities as a function of the probe sphere radius for representative values of $\epsilon_{\rm HH}$ and $\epsilon_{\rm HC}$. In \SIFig{fig:density}(a-c) we consider values within the droplet and absorbing droplet regimes, with panel (a) showing the number of proteins within the probe sphere $N_{\rm ps}$ as a function of radius. In both cases we observe ${N_{\rm ps}\sim r^3}$ for $r$ smaller than the droplet radius, as expected for a spherical droplet with a uniform density. \SIFig{fig:density}(b) shows the \textit{local} protein density, calculated as
\[
\rho_{\rm shell}(r)=\frac{N_{\rm shell}(r)}{4\pi r^2 dr} 
\]
where $N_{\rm shell}(r)$ is the number of proteins within the spherical shell of width $dr=0.3\sigma$ and radius $r$. We find that although $\rho_{\rm shell}(r)$ is rather noisy, it shows a clear drop to zero when $r$ reaches the droplet radius. In \SIFig{fig:density}(c) we plot the overall density within the probe sphere $\rho_{\rm ps}=N_{\rm ps}/V_{\rm ps}$ as a function of $r$, where ${N_{\rm ps}(r)=\int_0^{r} N_{\rm shell}(r') dr'}$ and $V_{\rm ps}=(4/3)\pi r^3$. We find that $\rho_{\rm ps}(r)$ is initially approximately constant with $r$, but at larger values it decreases towards the overall protein density ($\rho = 0.0233~\sigma^{-3}$). We use the plot in \SIFig{fig:density}(b) to obtain a lower bound for the droplet radius $r_{\rm in}$ (dashed line); then to obtain an estimate of the density in the protein rich phase, $\rho_{\rm HD}$, we fit a horizontal line to the plot in \SIFig{fig:density}(c) in the range $\sigma < r < r_{\rm in}$. Similarly, from \SIFig{fig:density}(b) we can also identify an outer radius, $r_{\rm out}$ (dotted line) which encompasses the entire droplet, plus any interface region or deviation due to the droplet not being exactly spherical. The density in the protein poor phase can then be estimated as $\rho_{\rm LD}=(N-N_{\rm ps}(r_{\rm out}))/(V_{\rm box}-V_{\rm ps})$, where $V_{\rm box}$ is the volume of the simulation box and $N=1000$ is the total number of proteins. We found that (except in the mixed regime) the choice $r_{\rm in}\!=\!4\sigma$, $r_{\rm out}\!=\!9\sigma$ was appropriate for all energy parameters ($\rho_{\rm ps}$ is approximately constant for $r \leq r_{\rm in}$ and scales as $r^{-3}$ for $r \geq r_{\rm out}$); we therefore used these values throughout. 

\SIFigures{fig:density}(d-f) show similar plots as detailed above, but comparing parameter values from the mixed or absorbing droplet regime. We note that the probing sphere procedure does not make sense in the mixed phase, where clusters of between 4 and 5 HP1s do form but only transiently: a different transient `largest cluster' is identified at each time point. We therefore do not consider the mixed regime results further. The $\epsilon_{\rm HH}\!=\!4k_BT$, $\epsilon_{\rm HC}\!=\!16k_BT$ case is in the absorbing droplet regime, and we note that the $\rho_{\rm ps}(r)$ curve in \SIFig{fig:density}(e) shows a slower drop off with $r$ at the droplet radius than that in \SIFig{fig:density}(b). The reason for this is that an absorbing droplet a significant fraction of the polymer extends out from the drop, and can be bound by coating proteins; this leads to a broader interface region. \SIFigures{fig:density}(g-i) show plots comparing the mixed and coating regimes. In the latter case the $\rho_{\rm ps}(r)$ curve is even broader; nevertheless, we can still estimate $\rho_{\rm HD}$ and $\rho_{\rm LD}$ using the above procedure.

For a given value of $\epsilon_{\rm HC}$, we can plot values of $\rho_{\rm HD}$ and $\rho_{\rm LD}$ on the $\rho$-$k_BT/\epsilon_{\rm HH}$ plane to obtain the more conventional phase diagram used for phase separating systems. Such plots are shown in \SIFigs{fig:denPhaseDiag}(a-b). By only including points for parameters where $\phi_{\rm sep}>0.6$ we can identify parts of the boundary between droplet and non-droplet phases [\SIFig{fig:denPhaseDiag}(a)]. By reducing this threshold to $\phi_{\rm sep}>0.2$ [\SIFig{fig:denPhaseDiag}(b)], we can also estimate boundaries for the coating regime, where we observe a partial phase separation (local increase in protein density due to coating the polymer as discussed in the main text). This allows us to sketch out the phase diagram for small and large values of $\epsilon_{\rm HC}$ in \mainFig{1(c)} in the main text, reproduced in \SIFig{fig:denPhaseDiag}(c) for completeness. As detailed in the main text and in \SIapp{sec:varyN} below, there is also a region [shaded bar in \SIFig{fig:denPhaseDiag}(c)] where a droplet only forms due to the presence of the polymer. Here the density of proteins within the two phases varies with the protein concentration.

\section{Hysteresis in the `droplet'--`absorbing droplet' transition}\label{sec:hyst}

In \mainFigs{3} and \mainFigJustNumber{4} of the main text we showed that there is an abrupt change in quantities such as the total fraction of proteins bound to the polymer $f_{\rm tot}$, the fraction of polymer beads bound by proteins $f_c$, and the polymer radius of gyration $R_g$, as $\epsilon_{\rm HC}$ is increased and the system moves from the droplet to the absorbing droplet regime. This hints that there may be a first order phase transition in the thermodynamic limit. To elucidate this further, we performed simulations where we slowly vary the parameter values in time and looked for evidence of hysteresis. In \mainFig{3(c)} in the main text we show a hysteresis loop for the fraction of proteins bound to the polymer (in total and in different modes) as $\epsilon_{\rm HC}$ is slowly increased from $8k_BT$ to $14k_BT$ before being decreased again.

To obtain this we ran $12$ independent repeat simulations for ${4\!\times\!10^4\tau}$, each starting from a different equilibrium configuration for $\epsilon_{\rm HH}=6k_BT$ and $\epsilon_{\rm HC}=8k_BT$ (droplet regime). For the first ${2\!\times\!10^4\tau}$ of each simulation $\epsilon_{\rm HC}$ is increased by an increment of ${3\!\times\!10^{-2}k_BT}$ every $10^2\tau$, until it reaches $\epsilon_{\rm HC}=14k_BT$ (the absorbing droplet regime). Then, over the second ${2\!\times\!10^4\tau}$ of the simulation $\epsilon_{\rm HC}$ is reduced in the same fashion (until $\epsilon_{\rm HC}=8k_BT$). We keep the protein-protein interaction energy constant throughout at $\epsilon_{\rm HH}=6k_BT$. For this intermediate value the droplet is highly dynamic (proteins often change their neighbours and there is relatively fast exchange of proteins between the droplet and the surrounding low density region, see \mainFig{6} in the main text); one would expect a slower response for larger $\epsilon_{\rm HH}$.

In \SIFig{fig:HyRg} we show how $R_g$ varies during the same simulations alongside the hysteresis plots for the fraction of proteins binding in different modes. This shows that there is also hysteresis in terms of the polymer configuration, as it retains memory of its previous state for a significantly long time after the system crosses the transition.

\section{Polymer distance maps}

To examine the polymer structure we can plot a `distance map' showing the mean 3D distance between every pair of polymer beads. \SIFigure{fig:CM}(a) shows maps from multivalent HP1 simulations representative of different regions of the phase diagram [\mainFig{2(c)} in the main text]. Each map is obtained from a single simulation of duration $5\!\times\!10^3\tau$. In the mixed and dense droplet regimes (not shown), distances between chromatin beads which are separated along the chain tend to be large and the configuration is dynamic. In the coating regime [\SIFig{fig:CM}(a)i], the chromatin is swollen, thus relative distances are again large, even though some distant regions along the chromatin chain can be connected by the very small fraction of HP1s which bind in the bridging mode. On the distance map this can be seen as a mixture of bright (large distance) and dark (short distance) regions.

In the absorbing droplet regime the distance map differs for different interaction energies. When $\epsilon_{\rm HC}$ is large but $\epsilon_{\rm HH}$ has an intermediate value [\SIFig{fig:CM}(a)ii], all of the chromatin is absorbed into the droplet in a crumpled configuration. Distances between chromatin beads which are separated along the chain tend to be short (dark colours in the map). If both interaction energies are large [\SIFig{fig:CM}(a)iii], the map shows mainly short distances but with a few bright stripes; the latter are the short regions of the chromatin which extend out of the droplet, and so tend to be further away from the rest of the polymer. \SIFigure{fig:CM}(a)iv shows the case of large $\epsilon_{\rm HH}$ but intermediate $\epsilon_{\rm HC}$. Here, quite long polymer segments loop out from the droplet: the map shows a coexistence of brighter regions (swollen chromatin) and darker regions (crumpled chromatin).

\SIFigure{fig:CM}(b) shows similar maps, but the standard deviation of the distance between beads is plotted instead of the mean. This gives a measure of how dynamic the polymer configuration is, with a larger standard deviation indicating more variation of the polymer bead separation in time (recall that each map is obtained from a single simulation run). \SIFigures{fig:CM}(b)iii-iv show that some regions have a strikingly low standard deviation (black); these corresponding to regions absorbed within the protein droplet. This implies little variation in polymer bead separations within the droplet (slow dynamics), with large variation of separation in the protruding loops (faster dynamics). For the parameters used in \SIFig{fig:CM}(b)ii, a fully absorbing (no protruding loops) droplet forms; here the variation in distances is more uniform and in the intermediate range, implying that the polymer is mobile within the droplet.

\section{Varying protein density}\label{sec:varyN}

As detailed in the main text, we performed simulations with different numbers of multivalent HP1s at two pairs of $\epsilon_{\rm HH}$,$\epsilon_{\rm HC}$ parameter values within the absorbing droplet regime. 

\SIFigure{fig:varyN} shows results for the case where ${\epsilon_{\rm HH}=6k_BT}$ and ${\epsilon_{\rm HC}=14k_BT}$; for these parameters the protein droplet would form even in the absence of chromatin. As can be observed from the snapshots in \SIFig{fig:varyN}(a), increasing the number of proteins leads to a larger droplet which absorbs a larger fraction of the polymer. We confirm quantitatively that the density of HP1s within the droplet is independent of the total number of HP1s (i.e., the overall density) using the probing sphere procedure detailed in \SIapp{sec:measureDensity} above. We plot the density within the probe sphere $\rho_{\rm ps}$ as a function of its radius $r$ in \SIFig{fig:varyN}(b), while the droplet radius for the three different values of $N$ is shown in \SIFig{fig:varyN}(c) [taken to be the position of the half maximum point in the $\rho_{\rm ps}(r)$ curve]. From this we see that when HP1-HP1 attraction drives droplet formation the behaviour is consistent with standard (Model-B) phase separation ($\rho_{\rm HD}$ is independent of $N$ and droplet radius $R_d$ increases as $N^{1/3}$).

\SIFigures{fig:varyN}(d-h) reveal a complicated relationship between the droplet size and chromatin absorption/compaction. This stems from the balance between the energetic gain which arises from HP1s binding chromatin, the entropic loss due to HP1 bound chromatin being confined to the volume of the droplet, and any energetic loss due to HP1-HP1 `bonds' being broken to accommodate HP1-chromatin `bonds'. 
For the $N=6000$ case the polymer is completely absorbed within the droplet, and we note that the volume which the polymer coil occupies is significantly smaller than the volume of the droplet. That is to say, the polymer is compacted to a greater extent that it would be due to simply being confined within the droplet. This is clear if one considers the ratio $R_g/R_d$, which steadily decreases with $N$ [\SIFig{fig:varyN}(e)]. To understand this, we varied the interaction strengths $\epsilon_{\rm HH}$ and $\epsilon_{\rm HC}$ by a small amount (such that we stay in the same regime) and observed the effect on the polymer radius of gyration (data not shown). We found that increasing $\epsilon_{\rm HH}$ led to greater compaction of the polymer. This is consistent with expectations if we consider the protein droplet to be an effective solvent within which the polymer is dissolved (e.g., as considered in Flory-Huggins theory).
On the other hand, increasing $\epsilon_{\rm HC}$ also led to greater compaction of the polymer; in the Flory-Huggins theory, increasing polymer-solvent attraction leads to \textit{swelling} of a polymer. Clearly the ability of our model HP1s to form bridges means that treating the droplet as a solvent gives an incomplete picture. 

\SIFigure{fig:varyN2} shows some additional results for the ${\epsilon_{\rm HH}=4k_BT}$, ${\epsilon_{\rm HC}=20k_BT}$ case (as in \mainFig{4} in the main text). This is the absorbing droplet regime, but here the droplet can only form in the presence of chromatin. In \SIFig{fig:varyN2}(a) we again consider a probing sphere of radius $r$ and plot $\rho_{\rm ps}(r)$, from which the droplet density $\rho_{\rm HD}$ is obtained (following the scheme described in \SIapp{sec:measureDensity}). \SIFigure{fig:varyN2}(b) shows how the droplet radius and polymer radius of gyration varies with $N$, as in \mainFig{4(g)} in the main text, but here on a linear rather than logarithmic scale. As above, we find that the ratio $R_g/R_d$ decreases as $N$ increases [\SIFig{fig:varyN2}(d)], but reaches a plateau as the growth of the droplet radius with $N$ slows.

\section{Droplet fractal dimension in the multivalent and limited valence models.}\label{sec:fractal}

In the snapshots of the limited valance model shown in \mainFig{5} in the main text we observe irregularly shaped protein clusters with a structure strikingly different to the spherical droplets formed by the multivalent model. To quantify this difference, here we estimate the fractal dimension $D_f$ of the clusters. In simulations, the fractal dimension of a cluster, e.g. from a diffusion limited cluster aggregation (DLCA) process~\cite{jungblut2019diffusion}, is typically obtained from a scatter plot of the cluster mass (or number or particles) \textit{versus} radius. For spherical clusters one would expect a scaling $R\sim N^{1/3}$, while fractal clusters give $R\sim N^{1/D_f}$ where $D_f<3$ for a 3D system.

In our simulations, we typically observe a single or small number of clusters, meaning it is difficult to obtain enough measurements to determine $D_f$. Another common method is to extract the fractal dimension from the structure factor $S(q)$~\cite{WU201341}, but this is again difficult to obtain from our simulations of a small number of clusters in a confined geometry. To estimate $D_f$, we instead consider smaller regions of the clusters, or ``sub-clusters'', measuring their mass and radius of gyration. We use the following scheme: we consider the $i$th HP1 together with all of its bound neighbours (defined as any HP1 whose centre of mass is within $1.1\sigma$ of HP1 $i$; different threshold values do not significantly alter the result). We denote this set of proteins a level 1 sub-cluster, and record the number of proteins $M$ and radius of gyration $R_g$ associated with this set. This is repeated for all HP1s in the system. We then consider level 2 sub-clusters, consisting of HP1 $i$, its bound neighbours, and all of the bound neighbours of neighbours; again, we record $M$ and $R_g$ for $i=1\dots N$. Level 3 sub-clusters include neighbours of neighbours of neighbours, etc. We continue increasing the level until there are no further unique sub-clusters, taking care not to double count. \SIFigure{fig:fractal} shows plots of $R_g$ against $M$ for all possible sub-clusters; each point represents the mean $R_g$ of all sub-clusters (of any level) with a given number of HP1s $M$.

For the multivalent model, for all parameters where there are protein droplets, the sub-cluster plots are roughly linear on a log-log scale, and have similar slope. A linear fit to a function $R_g=a M^{1/D_f}$ gives $D_f\approx 3.2$, close to the expected $D_f=3$ for spherical droplets. 

For the limited valence model, we find that sub-cluster plots are not always linear over the whole curve, and the exponent depends on the parameters. For large $\epsilon_{\rm HH}=20k_BT$ the plots are roughly linear with fractal dimension $D_f\approx 2.5$, which is insensitive to the value of $\epsilon_{\rm HC}$. This is close to the value $D_f=2$ observed in simulations of patchy particles~\cite{Audus2018}. For smaller $\epsilon_{\rm HH}=12k_BT$ there is not a single power law relationship between $R_g$ and $M$, but for large clusters $D_f\approx 3$. 
The reason for this difference is likely due to the difference in the protein dynamics. For $\epsilon_{\rm HH}=12k_BT$, protein in clusters can dynamically rearrange to satisfy the maximum number of bonds, tending to adopt more space-filling shapes; at larger $\epsilon_{\rm HH}$, HP1-HP1 bonds persist for long times, leading to dynamically arrested fractal clusters.

A variation on the above scheme, where neighbours are determined by considering actual interactions between NTE and hinge beads, gives similar $D_f$ values. Using the radius of the smallest cluster enclosing sphere instead of the radius of gyration gives slightly smaller $D_f$ values, but with a similar difference between the two models.

\section{Limited valence model: alternative quenching and additional figures.}

As noted above and in the main text, the limited valence model behaves similarly to patchy particles in that the system can adopt long long lived metastable states with multiple fractal clusters (including ``closed loops'' where all bonds are satisfied). The observed structures therefore depend on the initial condition or the quenching procedure used. To highlight this, in \SIFig{fig:snapalt}(a-b) we show configurations obtained with two different quenches. In \SIFig{fig:snapalt}(a), after stating from an equilibrium configuration for $\epsilon_{\rm HH},\epsilon_{\rm HC}=0$, first the HP1-chromatin attraction was switched on, then later the HP1-HP1 attraction was switched on. This procedure generated structures where most of the proteins were associated with the polymer, and were spread roughly uniformly along it; a few small (closed loop) clusters were not associated with the polymer. In \SIFig{fig:snapalt}(b), first the HP1-HP1 attraction was switched on, and then later HP1-chromatin attractions were switched on. In this case the proteins tend to sit in larger clumps associates with smaller sections of the polymer; therefore, much larger polymer regions are left without proteins bound. This latter morphology arises because the large HP1 clusters form first, only later becoming associated with the polymer. In all other limited valence simulations in this work we switched on both interactions at the same time. To demonstrate that the limited valence HP1 can form a gel, we also performed a simulation with periodic boundaries and a smaller box size (higher HP1 density); a snapshot is shown in  \SIFig{fig:snapalt}(c).

In \SIFigs{fig:ph_diag_LV}\textcolor{blue}{-S14} we present some additional measurements for the limited valence model HP1s. In \SIFig{fig:ph_diag_LV} we show the limited valence model phase diagram, drawing the crossover (or transition) lines between different regimes (or phases) in the same way as the multivalent case. We set the separation between the mixed and the fractal/absorbing clusters phases (orange line) where $\phi_{\rm sep} \approx 0.5$. As in the multivalent model, within the mixed regime $\phi_{\rm sep}$ is approximately independent of  $\epsilon_{\rm HC}$; we therefore define the coating regime as where $\phi_{\rm sep}$ starts to increase with $\epsilon_{\rm HC}$ (dashed green line). Interestingly, within the mixed regime $\phi_{\rm sep}$ shows more of a dependence on $\epsilon_{\rm HH}$ for the limited valence model than it did for the multivalent HP1s. Again similar to the multivalent case, for larger $\epsilon_{\rm HH}$ the absorbing clusters phase is defined as where the fraction of HP1s bound to chromatin $f_{\rm tot}\geq0.5$ [see \SIFig{fig:LVmodes}].
In \SIFig{fig:Rg} the fraction of chromatin beads bound to HP1s and the radius of gyration are shown. Note the non-monotonic behaviour of $f_c$ and $R_g$ is similar to the multivalent case.

\section{Protein Dynamics}\label{sec:dynamics}

As detailed in the main text, to quantify protein dynamics we consider the bond-bond correlation function $\nu_{\rm HH}(\Delta t)$, which measures the proportion of HP1s which retain the same interaction partners after a time interval $\Delta t$. \SIFigures{fig:dynamics}(a)i-iii show example plots of $\nu_{\rm HH}(\Delta t)$ for different values of $\epsilon_{\rm HH}$ and $\epsilon_{\rm HC}$ for the multivalent protein model. Clearly, these do not show a simple exponential decay (there is not a straight line on a log-linear plot); for small interaction energies $\nu_{\rm HH}$ decays quickly to very small values, while for large energies there is a fast initial decay followed by a much slower decrease. For cases where $\epsilon_{\rm HH}$ and $\epsilon_{\rm HC}$ are both large the $\nu_{\rm HH}$ curve almost plateaus at long time intervals [\SIFig{fig:dynamics}(a)iii]. Similar curves are shown in \SIFigs{fig:dynamics}(b)i-iii for the limited valence HP1 model. 

The non-exponential form of the $\nu_{\rm HH}(\Delta t)$ curves suggest that there are multiple time scales involved in this decorrelation. For example specific hinge-NTE bonds might break and form on short time scales, while repositioning of an HP1 with respect to its neighbours might take longer. We tried fitting several functional forms to the $\nu_{\rm HH}(\Delta t)$, finding a sum of three exponentials to consistently give good fits:
\begin{equation}\label{eq:3times}
\nu_{\rm HH}(t) = \frac{\mathrm{e}^{-t/\tau_0}+a_1 \mathrm{e}^{-t/\tau_1}+a_2 \mathrm{e}^{\tau_2}}{1+a_1+a_2},
\end{equation}
with decay times $\tau_0$, $\tau_1$, $\tau_2$, and dimensionless constants $a_1$ and $a_2$. We do not assign any specific meaning to these times or constants, but with analogy to a single exponential, we identify the integral of $t\times\nu_{\rm HH}(t)$ over $t$ as the mean decorrelation time
\begin{eqnarray}
\tau_{\rm HH}&=&\int_0^{\infty} \! t \,\frac{\mathrm{e}^{-t/\tau_0}+a_1 \mathrm{e}^{-t/\tau_1}+a_2 \mathrm{e}^{\tau_2}}{1+a_1+a_2} \,\mathrm{d}t \nonumber \\
&=& \frac{\tau_0+a_1\tau_1+a_2\tau_2}{1+a_1+a_2}.\nonumber
\end{eqnarray}
It is also common to fit a decorrelation function to a stretched exponential ($\exp [-(t/\tau_0)^\beta]$ with $\beta<1$), but we found this only gave a good fit for a few specific interaction energies. 
Grey lines in \SIFigs{fig:dynamics}(a)i-iii and \SIFigs{fig:dynamics}(b)i-iii show fits to the function in Eq.~(\ref{eq:3times}). The fits are good except for parameters corresponding to the mixed phase [lower curves in \SIFigs{fig:dynamics}(a)i and (b)i], where $\nu_{\rm HH}(\Delta t)$ drops quickly to very small values and the error bars are large. 

In \SIFigs{fig:dynamics}(a)iv-v and \SIFigs{fig:dynamics}(b)iv-v we show how $\tau_{\rm HH}$ varies with the interaction energies. This is similar to \mainFigs{6(b-c)} in the main text, except here we use the effective (re-scaled) energies $E_{\rm HH}$ and $E_{\rm HC}$, as defined in \SIapp{sec:energycalib}; we note there is very little difference in the shape of the curves. \SIFigures{fig:dynamics}(a)vi and (b)vi show the variation of $\tau_{\rm HH}$ with both energies as heat maps, with the boundaries between the regimes overlaid; this highlights some similarities and differences between the two models. In both cases, as expected $\tau_{\rm HH}$ is very small in the mixed phase. For $\epsilon_{\rm HC}$ and $\epsilon_{\rm HH}$ just large enough to enter the droplet/cluster phase, in both models $\tau_{\rm HH}$ is of the order $5\times10^2\tau$, and the droplet/clusters are highly dynamic on the time scale of the simulations (despite their different morphologies). Both models show a roughly order of magnitude increase in $\tau_{\rm HH}$ as $\epsilon_{\rm HC}$ is increased and chromatin becomes absorbed. The models differ substantially only at higher $\epsilon_{\rm HH}$ values: in the limited valence case there is no longer a large slow-down as the polymer becomes absorbed. It is interesting to note that although the limited valence model displays clusters with a fractal/gel-like morphology usually associated with arrested dynamics, the bond-bond decorrelation time is similar to the multivalent case. In this sense both models display similarly ``arrested dynamics'', but while all configurations of the multivalent proteins look similar macroscopically (a spherical droplet), the limited valence fractal clusters have a macroscopic appearance which depends on their history. 


\balance